\documentclass[preprint]{emulateapj}

\shortauthors{Adams {\it et al.}}
\shorttitle{Tidal Features in Galaxy Clusters}

\begin{document}

\title{The Environmental Dependence of the Incidence of Galactic Tidal Features}

\author{Scott M. Adams$^{1,2}$, Dennis Zaritsky $^1$, David J. Sand$^{3,4}$, Melissa L. Graham$^{3,5}$, Chris Bildfell$^5$, Henk Hoekstra$^6$, \& Chris Pritchet$^5$}

\affil{$^1$ Steward Observatory, University of Arizona, Tucson, AZ, 85721, USA}

\affil{$^2$ Department of Astronomy, The Ohio State University, Columbus, OH, 43210, USA}

\affil{$^3$ Las Cumbres Observatory Global Telescope Network, 6740 Cortona Dr., Suite 102, Santa Barbara, CA, 93117, USA}

\affil{$^4$ Department of Physics, Broida Hall, University of California, Santa Barbara, CA, 93106}

\affil{$^5$ Department of Physics and Astronomy, University of Victoria, P.O.Box 3055, STN CSC, Victoria, BC, V8W 3P6, Canada}

\affil{$^6$ Leiden Observatory, Leiden University, Niels Bohrweg 2, NL-2333 CA
Leiden, The Netherlands}

\begin{abstract}

In a sample of 54 galaxy clusters $(0.04<z<0.15)$ containing 3551 early-type
galaxies suitable for study, we identify those with tidal features both
interactively and automatically. We find that $\sim$ 3\% have tidal features
that can be detected with data that reaches a $3 \sigma$ sensitivity limit of
26.5 mag arcsec$^{-2}$.  Regardless of the method used to classify tidal
features, or the fidelity imposed on such classifications, we find a deficit of
tidally disturbed galaxies with decreasing clustercentric radius that is most
pronounced inside of $\sim 0.5R_{200}$.  We cannot distinguish whether the trend
arises from an increasing likelihood of recent mergers with increasing
clustercentric radius or a decrease in the lifetime of tidal features with
decreasing clustercentric radius.  We find no evidence for a relationship
between local density and the incidence of tidal features, but our local density
measure has large uncertainties.  We find interesting behavior in the rate of
tidal features among cluster early-types as a function of clustercentric radius
and expect such results to provide constraints on the effect of the cluster
environment on the structure of galaxy halos, the build-up of the red sequence
of galaxies, and the origin of the intracluster stellar population.
\end{abstract}

\section{Introduction}

Galaxies and clusters of galaxies assemble hierarchically, through a long
progression of mergers and accretion. As one considers galaxy-scale dark matter
halos, baryonic physics becomes more integral and so the analytic formalism
developed to calculate the rate of dark matter halo evolution
\citep{ps,bond,lc93}, which is broadly confirmed by numerical simulations
\citep{lc94,sheth}, becomes less prescriptive. To remedy this shortcoming,
galaxy formation is now studied using hydrodynamic numerical simulations
\citep[e.g.,][]{springel,hoffman}, although the details of the assembly process
are sensitive both to baryonic physics below scales accessible in the
simulations (``sub-grid" physics) and to the complex history that occurs on
cosmological scales. Tests of such models include comparisons to the properties
of galaxy clusters, their constituent galaxies, and the generation of
intracluster stars \citep{puchwein}.

The rate of accretion events and mergers is a fundamental quantity that
hierarchical growth models must be able to predict accurately.  A proper
comparison to an empirical rate measurement is elusive \citep[see][for a
description of the current state of the field]{lotz}.  Efforts to measure this
rate, particularly as a function of redshift, have focused either on measuring
close pairs of galaxies \citep[a few examples include][]{carlberg, patton,
lefevre,kartaltepe,deravel}, or identifying galaxies that appear morphologically
to be the result of ongoing or recent merger events \citep[a few examples
include][]{abraham,lotz08,jogee,bridge,miskolczi} and results often disagree
among studies \citep{lotz}.  These measurements are difficult, even with the
superb angular resolution of the {\sl Hubble Space Telescope}, for high redshift
galaxies.  However, renewed focus on identifying tidal features at low redshift
has yielded some striking successes both in the increased incidence of detected
features \citep{tal} and their scale and morphologies \citep{md}. 

We aim to make a measurement of the major merger rate at low redshift, but as a
function of environment. A complication in practice is that the definition of a
major, vs a minor, merger is typically made on the basis of the mass ratio
between the two galaxies. Such a definition is difficult to apply empirically,
particularly in environments that may have affected the individual galaxies and
in cases where the two initial galaxies have merged into a single remnant. We
define a major merger here as one that results in the large-scale tidal features
we are searching for around luminous, early-type ($R^{1/4}$ profile) galaxies.
Although this approach evidently complicates comparisons to models, it should
not impact our primary goal of searching for patterns in the occurrence of tidal
features as a function of environment.
 
 \cite{dubinski} showed how the dark matter halo structure of the merging
galaxies can significantly alter the appearance of a major merger, thereby
demonstrating that measurements of major merger signatures, even at a single
redshift, could be a sensitive model diagnostic.  Beyond the purpose of
understanding hierarchical accretion better, we need to study the merger rate in
clusters and in the surrounding environments because these events contribute
stars to the ubiquitous intracluster stellar population
\citep{zibetti,gonzalez05}, which is an important component in estimates of the
baryon budget of groups and clusters \citep{gonzalez07,giodini,mcgee}, and
calculations of the chemical enrichment history of the intracluster medium
\citep{zaritsky,sivanandam}.  Furthermore, extending any study of galaxies
within clusters out to the surrounding environs is critical because differences
in their star formation histories, which could be related to interactions, begin
to appear at clustercentric radii of several Mpc \citep{lewis,gomez}. 

The principal observational challenge in identifying tidal features is that such
features are of low surface brightness and have irregular morphologies.  A
variety of techniques have been developed to identify merging galaxies and
mergers \citep[e.g.,][]{conselice,lotz08}, but the problem takes on different
forms depending on the redshift of the galaxies and the depth and uniformity, or
``flatness", of the data.  For galaxies in the local universe, \cite{colbert}
employed unsharp masking (subtracting a smoothed image of the galaxy from
itself), galaxy model division (dividing the galaxy image by a best-fit model
galaxy made from nested elliptical isophotes), and color mapping (division of
the galaxy image in one color band by the galaxy image in another color band) to
aid in the visual detection of tidal features and found that 41\% (9 of 22) of
analyzed isolated galaxies show tidal features, while only 8\% (1 of 12) of
group galaxies had such features.  While this result is certainly suggestive of
lower rates of tidal features in group environments than in the field, Poisson
statistics limits the significance of the discrepancy to less than $2\sigma$.
\cite{vandokkum} divided his galaxy images by a model fit and measured the mean
absolute deviation of the residuals to define a quantitative threshold for the
identification of tidally disturbed galaxies.  He found that 53\% of his sample
of 126 red, field galaxies show tidal features, while 71\% of the
bulge-dominated early-type subset of 86 galaxies appear to be tidally disturbed.
Using a similar method for a sample of nearby, luminous, elliptical galaxies,
\cite{tal} found that 50\% (5 of 10) of cluster galaxies, 62\% (13 of 18) of
poor-group galaxies, and 76\% (16 of 21) of isolated galaxies have tidal
features.  Similar to the results of \cite{colbert}, these also suggest that
tidal features are less common in cluster environments than in the field, but
again the significance of the result is limited by the sample size.
\cite{jan10} quantify the extent of tidal substructure in five elliptical
galaxies in the Virgo cluster by measuring the luminosity of the substructure
and find no obvious correlation between clustercentric radius and the amount of
substructure.  Larger samples are needed to resolve this issue.

Larger samples studied thus far are from field surveys.  \cite{bridge} visually
identify tidal tails in a sample of 27,000 galaxies over 2 square degrees of the
Canada-France-Hawaii Telescope Legacy Deep Survey (CFHTLS-Deep) to find that the
merger fraction of galaxies evolves from 4\% at $z \sim 0.3$ to 19\% at $z \sim
1$.  \cite{miskolczi} study a sample of 474 galaxies selected from the SDSS DR7
archive to find that at least 6\% of the galaxies have distinct tidal streams
and a total of 19\% show faint features.  No large samples have included
substantial numbers of galaxies in dense environments.

Deep, uniform imaging is a prerequisite for any such study. The incidence of
identified mergers among field galaxies increases from $<$ 10\% locally in
relatively shallow imaging \citep{miskolczi} to many tens of percent in deeper
imaging \citep{md, tal}.  For this and other reasons, it is neither surprising
nor incorrect that the apparent total rates from different studies vary widely.
The key is therefore to compare within a given study rather than across studies. 

We identify tidally disturbed galaxies in the largest existing sample of deep,
wide-field images of nearby galaxy clusters $(0.04 < z < 0.15)$. These come from
the Multi-Epoch Nearby Cluster Survey \citep[MENeaCS; ][]{sand2}, and this study
is part of a series using those data to address a series of science questions
including so far the incidence of intracluster supernovae \citep{sand1}, the
rate of SNe Ia in clusters \citep{sand2}, the evolution of the dwarf-to-giant
ratio \citep{bildfell}, and the rate of SNe II's in clusters \citep{graham}.
The larger sample size enables us to examine, for the first time, relative rates
of tidal features as a function of different measures of the global
environmental (projected clustercentric radius) and local environment (projected
local galaxy density). In \S2 we briefly present the data we use and discuss our
analysis of these data, including the selection of candidate cluster galaxies
and the calculation of the tidal parameter. In \S3 we present our results and
discuss whether any significant trends are identified. We present our
conclusions in \S4.  We adopt $H_{0}=71$ km s$^{-1}$ Mpc$^{-1}$,
$\Omega_{M}=0.27$, and $\Omega_{\Lambda}=0.73$ for conversions to physical
units.

\section{Data and Analysis}

\subsection{The Cluster Sample} The galaxy cluster images used for this study
are the deep stack images obtained by the Multi-Epoch Nearby Cluster Survey of
X-ray selected clusters \citep[MENeaCS;][]{sand2} with MegaCam on the
Canada-France-Hawaii Telescope \citep{boulade}.  The image stacks typically
contain 20 to 30 120-second exposures in the $g^\prime$ and $r^\prime$ filters
for each cluster, observed at a monthly cadence to pursue the supernovae science
goals \citep{sand1,sand2}.  The analysis is done on the deeper $r^\prime$ data.
The depth of the $r^\prime$ image stacks for point source detection corresponds
to limiting absolute magnitudes of $-$11 to $-$13 at the distance of the
clusters.  The $3\sigma$ limiting apparent surface brightness, derived from the
sky RMS, of 1 arcsec$^{2}$ patches in the deep stack images is approximately
26.5 mag arcsec$^{-2}$.  The faintest tidal features that we visually identify
are approximately this magnitude (see Fig. \ref{fig-visual_and_auto} and
\ref{fig-visual_or_auto}).  MegaCam features a large field of view ($\sim$1
degree$^{2}$) that enables us to search for tidal features from the central
regions out to $1-2R_{200}$ and study the radial and environmental dependence of
the merger rate.  Details on the cluster parameters and image reduction process
are given in \cite{sand1,sand2}.  The list of the 54 clusters analyzed is given
in Table \ref{tab:clusters}.  The cluster redshifts range from $0.04<z<0.15$,
with the redshifts taken from the NASA Extragalactic Database.

\begin{deluxetable*}{lrrrrrrr}
\tablewidth{0pt}
\tablecaption{Galaxy Clusters}
\tablehead{
\colhead{Cluster} &
\colhead{Redshift} &
\colhead{$L_{X,0.1-2.4 keV}$} &
\colhead{${M_{200}}$\tablenotemark{a}} &
\colhead{${R_{200}}$\tablenotemark{b}} & \colhead{$N_{gal}$} & \colhead{$N_{gal}$} & \colhead{$N_{gal}$}
 \\
\colhead{} & \colhead{} &
\colhead{($10^{44}$ ergs/s)} &
\colhead{($M_{\odot}$)} &
\colhead{(Mpc)} &
\colhead{Analyzed} & \colhead{Visual Sample} & \colhead{Cleaned-auto sample}
}
\startdata
Abell1033 & 0.126 & 5.12 & 7.28 & 1780  & 72 & 2 & 1 \\
Abell1068 & 0.138 & 5.94 & 8.13 & 1840  & 99 & 11 & 2 \\
Abell1132 & 0.136 & 6.76 & 7.44 & 1790  & 107 & 8 & 2 \\
Abell119 & 0.044 & 3.30 & 5.14 & 1630  & 32 & 0 & 3 \\
Abell1285 & 0.106 & 4.66 & 6.76 & 1750  & 93 & 15 & 4 \\
Abell133 & 0.057 & 2.85 & 4.78 & 1580  & 38 & 2 & 1 \\
Abell1348 & 0.119 & 3.85 & 5.94 & 1670  & 71 & 3 & 2 \\
Abell1361 & 0.117 & 4.95 & 7.09 & 1770  & 57 & 2 & 1 \\
Abell1413 & 0.143 & 10.83 & 12.45 & 2120  & 153 & 5 & 0 \\
Abell1650 & 0.084 & 5.66 & 9.18 & 1950  & 72 & 5 & 0 \\
Abell1651 & 0.085 & 6.92 & 10.48 & 2040  & 44 & 5 & 2 \\
Abell1781 & 0.062 & 3.79 & 5.73 & 1680  & 30 & 1 & 2 \\
Abell1927 & 0.095 & 2.30 & 4.08 & 1490  & 62 & 0 & 0 \\
Abell1991 & 0.059 & 1.42 & 2.86 & 1340  & 26 & 2 & 2 \\
Abell2029 & 0.077 & 17.44 & 16.57 & 2380  & 75 & 4 & 2 \\
Abell2033 & 0.082 & 2.55 & 4.38 & 1530  & 76 & 2 & 0 \\
Abell2050 & 0.118 & 2.63 & 4.54 & 1530  & 86 & 6 & 3 \\
Abell2055 & 0.102 & 3.80 & 5.83 & 1670  & 54 & 1 & 1 \\
Abell2064 & 0.108 & 2.96 & 4.92 & 1570  & 55 & 1 & 4 \\
Abell2069 & 0.116 & 3.45 & 5.49 & 1630  & 124 & 3 & 1 \\
Abell21 & 0.095 & 2.64 & 4.51 & 1540  & 43 & 2 & 4 \\
Abell2142 & 0.091 & 21.24 & 19.56 & 2510  & 130 & 1 & 2 \\
Abell2319 & 0.056 & 15.78 & 15.61 & 2350  & 24 & 3 & 2 \\
Abell2409 & 0.148 & 7.57 & 9.69 & 1950  & 88 & 0 & 0 \\
Abell2420 & 0.085 & 4.64 & 6.67 & 1760  & 64 & 0 & 3 \\
Abell2426 & 0.098 & 4.96 & 7.04 & 1780  & 88 & 0 & 1 \\
Abell2440 & 0.091 & 3.36 & 5.33 & 1630  & 63 & 0 & 0 \\
Abell2443 & 0.108 & 3.22 & 5.22 & 1600  & 46 & 0 & 1 \\
Abell2495 & 0.078 & 2.74 & 4.58 & 1550  & 37 & 0 & 1 \\
Abell2597 & 0.085 & 6.62 & 8.62 & 1910  & 39 & 0 & 3 \\
Abell2627 & 0.126 & 3.25 & 5.29 & 1600  & 80 & 0 & 0 \\
Abell2670 & 0.076 & 2.28 & 4.03 & 1490  & 51 & 0 & 3 \\
Abell2703 & 0.114 & 2.72 & 4.64 & 1540  & 86 & 1 & 2 \\
Abell399 & 0.072 & 7.06 & 8.15 & 1880  & 89 & 4 & 9 \\
Abell401 & 0.074 & 12.06 & 13.02 & 2200  & 82 & 5 & 5 \\
Abell553 & 0.066 & 1.83 & 3.43 & 1410  & 12 & 3 & 2 \\
Abell644 & 0.070 & 8.33 & 10.02 & 2020  & 23 & 2 & 2 \\
Abell646 & 0.129 & 4.94 & 6.47 & 1710  & 61 & 1 & 2 \\
Abell655 & 0.127 & 4.90 & 6.54 & 1720  & 97 & 1 & 0 \\
Abell7 & 0.106 & 4.52 & 6.61 & 1740  & 29 & 0 & 1 \\
Abell754 & 0.054 & 7.00 & 10.11 & 2040  & 47 & 3 & 3 \\
Abell763 & 0.085 & 2.27 & 4.03 & 1480  & 40 & 2 & 3 \\
Abell780 & 0.053 & 4.78 & 6.72 & 1780  & 22 & 1 & 0 \\
Abell795 & 0.136 & 5.70 & 7.89 & 1820  & 107 & 5 & 5 \\
Abell85 & 0.055 & 9.41 & 10.27 & 2050  & 47 & 3 & 2 \\
Abell961 & 0.124 & 3.12 & 5.13 & 1590  & 106 & 3 & 3 \\
Abell990 & 0.144 & 6.71 & 8.88 & 1890  & 147 & 2 & 4 \\
MKW3S & 0.045 & 3.45 & 4.09 & 1510  & 12 & 0 & 1 \\
ZwCl1023 & 0.143 & 4.71 & 6.92 & 1740  & 81 & 1 & 3 \\
ZwCl1215 & 0.075 & 5.17 & 7.27 & 1810  & 55 & 0 & 0 \\
\enddata
\tablecomments{Basic properties of clusters analyzed along with the total number of galaxies and number in the visual and cleaned-auto samples of disturbed galaxies for each cluster.}
\tablenotetext{a}{$M_{200}$ derived from $L_{X,0.1-2.4 keV}$ using relation found by \cite{reiprich}}
\tablenotetext{b}{$R_{200}$ found from correspondence with $M_{200}$}
\label{tab:clusters}
\end{deluxetable*}

\subsection{Selecting Elliptical Cluster Galaxies}

We use SExtractor, a source detection algorithm \citep{bertin}, in two-image
mode to measure the magnitudes and colors of all galaxies in the images.  The
galaxy clusters are relatively nearby ($z < 0.15$) and we can only detect tidal
features in galaxies of relatively large angular extent (many arcsec), so we
select for further study only sources that have a minimum area of 60 pixels that
are each at least 6$\sigma$ above the background and have a stellarity parameter
of less than 0.05.  Using that output, we create color magnitude diagrams for
each cluster using the absolute magnitudes derived from SExtractor's
AUTO$_{-}$MAG parameter, which uses an aperture radius of $2.5 r_{Kron}$
\citep{kron}, and the color using the APER$_{-}$MAG parameter, which finds the
magnitude within a small aperture we set to 1.9" in diameter in order to
maintain high S/N and limit contributions from neighboring sources. 

\begin{figure}[]
\begin{center}
\plotone{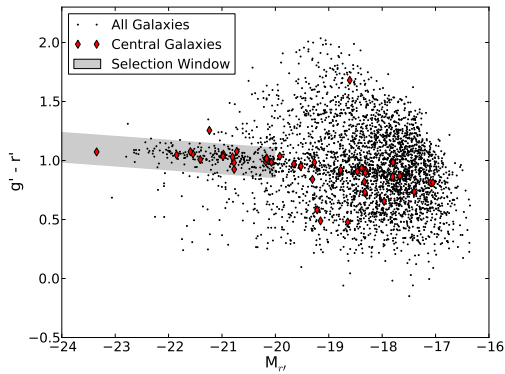}
\caption{Sample color magnitude diagram (Abell 21).  The black dots represent all galaxies meeting our SExtractor selection parameters.  The red diamonds are galaxies located within 1.5' of the cluster center used to originally define the location of the cluster red sequence.  Galaxies within the shaded region are brighter than our minimum magnitude and are identified as red sequence members of the cluster.  These are the galaxies in which we search for tidal features.\label{fig-cm}}
\end{center}
\end{figure}

To identify a sample of galaxies with a high probability of being at the cluster
redshift and to avoid star-forming, late-type galaxies that can have significant
internal structure, we select red sequence galaxies for further study.  To do
this, we first visually examine the location of galaxies in the color-magnitude
diagram that are within 1.5 arcmin of the cluster center. Using these galaxies,
we define the color normalization of the red sequence and adopt an
empirically-determined slope of $-$0.033. The red sequence slopes of clusters at
a given z show little if any scatter \citep{ellis,stanford, gilbank}.  We select
galaxies with colors within $\pm  0.13$ of the mean relation for further study.
This estimate of the color scatter is probably somewhat conservative (for
example, \cite{bildfell} select galaxies within $\pm 0.2$ of the red sequence),
but it minimizes contamination, which for this study is more important than
completeness.  A tight color selection could result in a bias if interacting
galaxies are bluer than the mean, but (as we find later) they are not.  We do
not explicitly apply k-corrections or evolution corrections, but they are
inherent in our red-sequence color selection.  We repeat the procedure of
visually identifying the optimal zero point for each cluster, keeping the slope
and thickness of the color-magnitude band constant.  We then apply these cuts to
the entire image, keeping only galaxies with $M_{r}<-20$ (where $M_{r}$ is
simply the sum of the apparent magnitude and the distance modulus) at the
cluster redshift (see Fig. \ref{fig-cm} for a sample color-magnitude diagram
and galaxy selection) to ensure that the candidate galaxies are sufficiently
luminous for our tidal analysis. We will discuss estimates of the interloper
fraction in \S3.2.

\begin{figure}[]
\begin{center}
\plotone{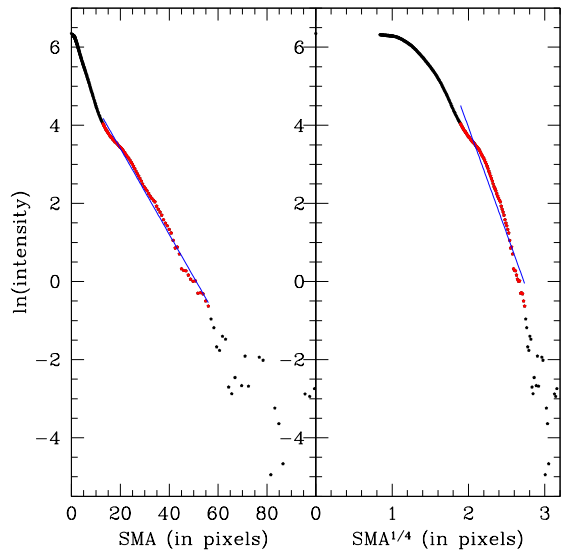}
\caption{Sample luminosity profile fit used to identify a spiral galaxy.  The linear least-squares fit of the ln(intensity) versus semi-major axis (left panel) has a smaller $\chi^{2}$ profile than the least-squares fit of the ln(intensity) versus semi-major axis$^{1/4}$.  The portions of the luminosity profile used for the best fits are shown in red, the portions excluded are shown in black.  The blue lines are the best fit linear regression lines.\label{fig-spiral_fit}}

\end{center}
\end{figure}

We focus our study on elliptical galaxies because their surface brightness
distributions can be modeled more easily than those of spirals and because disk
galaxies are believed to not have been involved in major mergers since moderate
redshifts unless both progenitors were gas rich, a condition that will be rare
in these high density environments \citep{robertson}.  To limit our sample to
elliptical galaxies, we reject galaxies with ellipticity greater than 0.5 and we
discriminate based on the shape of the surface brightness profile.  To do the
profile discrimination, we select the radial region of the galaxy's surface
brightness profile in which the intensity is less than 10\% of the galaxy's peak
intensity (to avoid the exponential bulge of spirals) and greater than $1\sigma$
above the background sky level.  We reject galaxies that have a lower $\chi^{2}$
for the weighted least squares linear fit of $\ln$(intensity) vs. radius than of
$\ln$(intensity) vs $r^{1/4}$ (Fig. \ref{fig-spiral_fit}).  Out of the initial
sample of 11,904 red sequence galaxies, 4526 were flagged as disks due to their
surface brightness profiles and 2322 galaxies were flagged as disks due to their
ellipticity, with a union of the two criteria resulting in a total of 5486
galaxies being dropped from the initial sample. We found with visual examination
that only $ \sim 3\%$ of galaxies classified as ellipticals had obvious spiral
structure (these are later discarded from the final sample).  It is more
difficult to estimate the fraction of galaxies identified as disks that are
actually ellipticals due to the similar morphology of E and S0 galaxies, but the
vast majority of the galaxies identified as disks did visually appear to be
correctly flagged.  There were some instances ($<1$\% of all galaxies) where
obvious mergers were classified as disks, but these tended to be galaxies that
still had an obvious disk component and/or lacked a clear nucleus.

For the subsequent analysis, we extract image regions of width corresponding to
seven times each galaxy's diameter (as determined by SExtractor). The sizes of
these images range from 46 arcmin$^{2}$ for the largest central galaxies to 500
arcsec$^{2}$ for the smallest galaxies that meet our thresholds, with the median
size being 1 arcmin$^{2}$.  The angular sizes translate into 850$\times$850 kpc
to 50$\times$50 kpc with the median area being 105$\times$105 kpc.  The sizes of
the images were chosen to ensure that any possible extended tidal tails would be
included in the analysis and that in almost all cases there is sufficient area
to accurately define a background sky level.  Visual examination did not reveal
any cases where tidal features began beyond the analyzed image region.  Though
there are a few cases where large tidal features continue beyond the analyzed
image region, these galaxies are still correctly identified as tidally-disturbed
by both the visual and automated selections.

\subsection{Calculating the Tidal Parameter}

In studies of mergers and merger remnants there is always a tension between
visual and automated identification. Visual approaches have the advantage that
they harness the tremendous pattern recognition skills that the human eye has
evolved, but suffer from non-uniformity among different practitioners, and
poorly defined criteria. Automated approaches are repeatable and well-defined,
the latter being particularly important for comparisons to models, but are
likely to be less powerful in selecting varying morphologies within this
ill-defined class of objects. We resolve this conflict by applying both
approaches.

\begin{figure}[]
\plotone{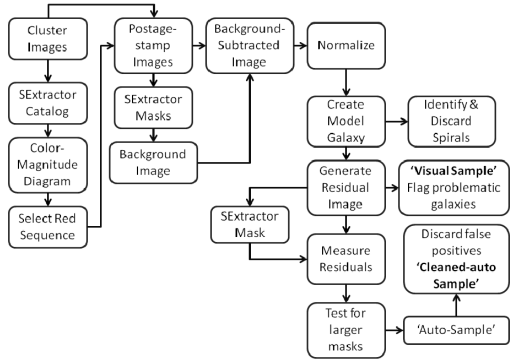}
\caption{Flow chart showing process by which samples of tidally-disturbed galaxies are identified.\label{fig-flowchart}}
\end{figure}

To objectively determine whether a galaxy is tidally-disturbed we follow the
general procedure presented by \cite{tal} (shown in Fig. \ref{fig-flowchart}) to
calculate a tidal parameter from the average residual of the galaxy image
divided by its best fit model. First a background gradient is fit and subtracted
from each galaxy image using the BACKGROUND generated by SExtractor with
BACK$_{-}$SIZE = 64 and BACK$_{-}$FILTERSIZE = 3, in order to minimize large
scale gradients across the galaxy images caused by intracluster light or nearby
bright galaxies.  The BACKGROUND map generated by SExtractor is a bicubic spline
interpolation over the sigma-clipped pixels in the area of the BACK$_{-}$SIZE
parameter median-filtered over the BACK$_{-}$FILTERSIZE number of BACK$_{-}$SIZE
areas.  The BACK$_{-}$SIZE parameter is chosen for our pixel scale
(0.187"/pixel) to be larger than the scales of typical tidal features.

The tidal parameter analysis is very sensitive to unmasked sources, which are
easily confused with tidal matter. Pixels with known objects are masked using
the SExtractor object catalog.  To optimize masking of background sources while
minimizing masking of real tidal features, we create our masks from the union of
two different masking catalogs.  One catalog targets large galaxies using
DETECTMIN$_{-}$AREA $=$ 5, DETECT$_{-}$THRESH $=$ 1.8, and DEBLEND$_{-}$MINCONT
$=$ 0.001 to only mask very significant sources well removed from the subject
galaxy.  These parameters require DETECTMIN$_{-}$AREA number of neighboring
pixels to be DETECT$_{-}$THRESH sigma above the background in order to be
detected as an object.  The DEBLEND$_{-}$MINCONT parameter determines if a
cluster of pixels above the detection threshold should be divided into multiple
sources, and has been tuned here to differentiate only very distinct clumps as
separate objects.  A second, more aggressive catalog targeting stars is
generated by only identifying objects with CLASS$\_$STAR less than 0.05 with
DETECT$_{-}$THRESH $=$ 1.4, and DEBLEND$_{-}$MINCONT $=$ 0.0001 (a lower
deblending contrast to more readily identify stars blended with the target
galaxy).  We also create a background sky frame to be used in a noise correction
term (described below) using a third, very agressive catalog generated with
DETECT$_{-}$THRESH $=$ 0.8, and DEBLEND$_{-}$MINCONT $=$ 0.001 that masks all
sources (including the target galaxy and associated tidal features).  Unmasked
pixels are randomly chosen to replace the masked pixels in the creation of a
background image in order to preserve the noise characteristics of the original
image.

We create a best-fit model for each galaxy using the IRAF tasks \textit{ellipse}
and \textit{bmodel}.  We fix the isophotal center but allow for variable
position angle and ellipticity as a function of semi-major axis to generate
best-fit isophotes out to 80\% of the postage stamp image dimensions, and then
create the corresponding model galaxy.  The model image covers the entire
postage stamp area, with pixel values set to zero beyond the area covered by the
best fit model.  Pixels whose model value exceeds 0.01 times the peak value at
the galaxy center are masked so that the analysis focuses on any extended tidal
features rather than residuals near the galaxy nucleus that may arise from
slight misplacement of the galaxy center in the model.  The masked galaxy image
is then divided by the model image.  The residual image is median filtered with
a 5$\times$5 pixel kernel in order to facilitate visual identification of tidal
features.  We further supplement our pixel mask by running SExtractor on the
smoothed residual image with DETECTMIN$_{-}$AREA $=$ 5, DETECT$_{-}$THRESH $=$
5, and DEBLEND$_{-}$MINCONT $=$ 0.0001 primarily in order to mask faint galaxies
close to the primary galaxy.  The final mask for a typical galaxy includes
5-20\% of the pixels in the area over which the model is computed.  We discard
all galaxies that have over 50\% of the image area masked.  This should not bias
the results unless real tidal features are being masked (which we visually
verified is very uncommon, with only a few cases out of several thousand
galaxies). We also confirm with a control sample that the masking is not
responsible for the radial trend we discuss below (see \S3.2).  With the
enhanced mask we repeat the process of creating a model galaxy and smoothed
residual image.  

We examine all galaxy and residual images for quality and calculate the tidal
parameter.  We visually inspect and discard the galaxy and corresponding
residual images that are contaminated with  unmasked diffraction spikes from
bright stars.  We also flag galaxies that have visible tidal structure. This
latter sample is referred to as our visual sample. We then generate an
auto-detected sample using a tidal parameter that is calculated by finding the
mean of the absolute value of the final residual image: \begin{equation}
T_{galaxy}=\overline{\bigg| \frac{I_{x,y}}{M_{x,y}}-1\bigg|} \end{equation}
where I$_{x,y}$ is the pixel value at x,y of the object frame and M$_{x,y}$ is
the pixel value at x,y of the model frame.  A background-noise correction frame
is created by adding the background sky image created earlier to the model image
and then dividing by the model.  The resulting frame is then median filtered and
a tidal parameter is recalculated for the model plus noise frame.  The final
tidal parameter, corrected for noise, is then found as follows: \begin{equation}
T_{c}=\sqrt{T_{galaxy}^{2}-T_{model}^{2}} \end{equation}

This tidal parameter is particularly sensitive to deviations when the model
values are small.  Because our models often go to zero toward the image edges,
we find that we increase the relative stability of our tidal parameters by 
first dividing the galaxy images by their mean background counts so that all
images have a pedestal of 1 count.

\begin{figure}
  \centerline{
    \includegraphics[width=4cm, angle=0]{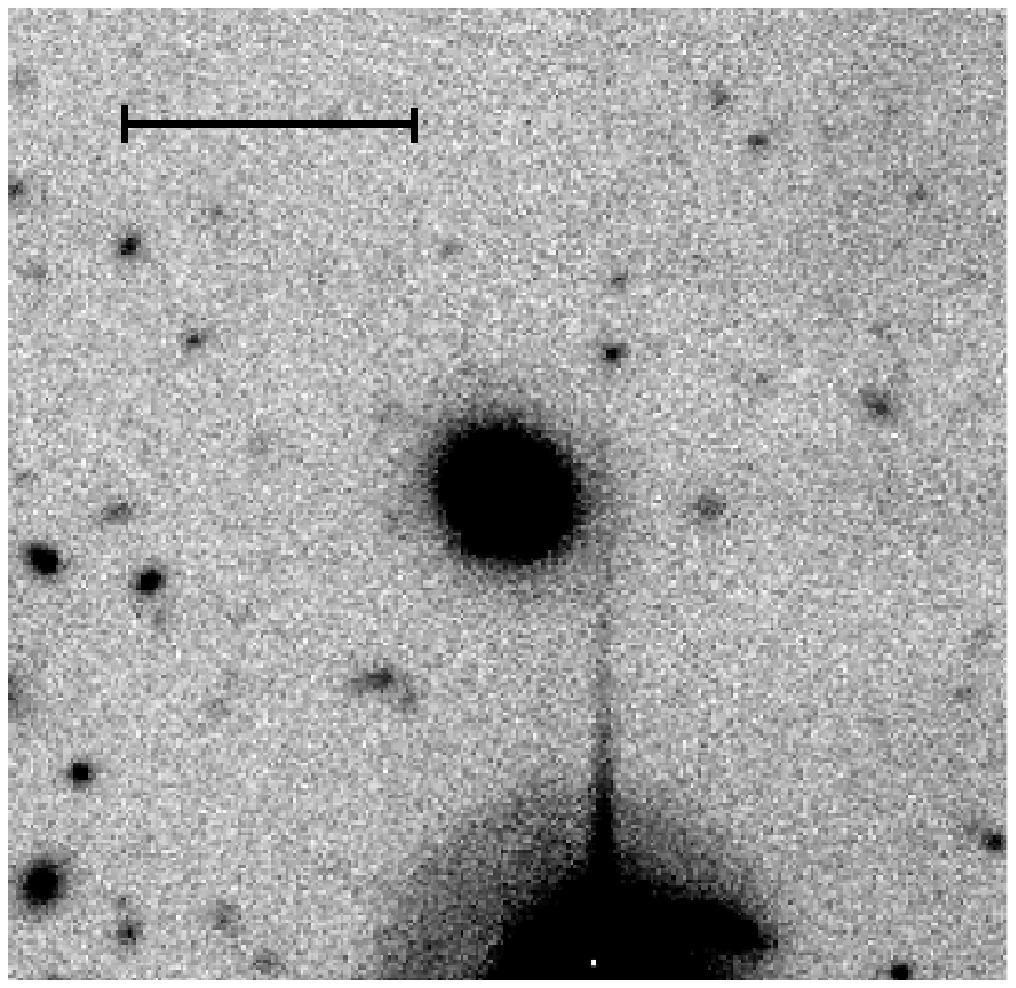}%
    \includegraphics[width=4cm, angle=0]{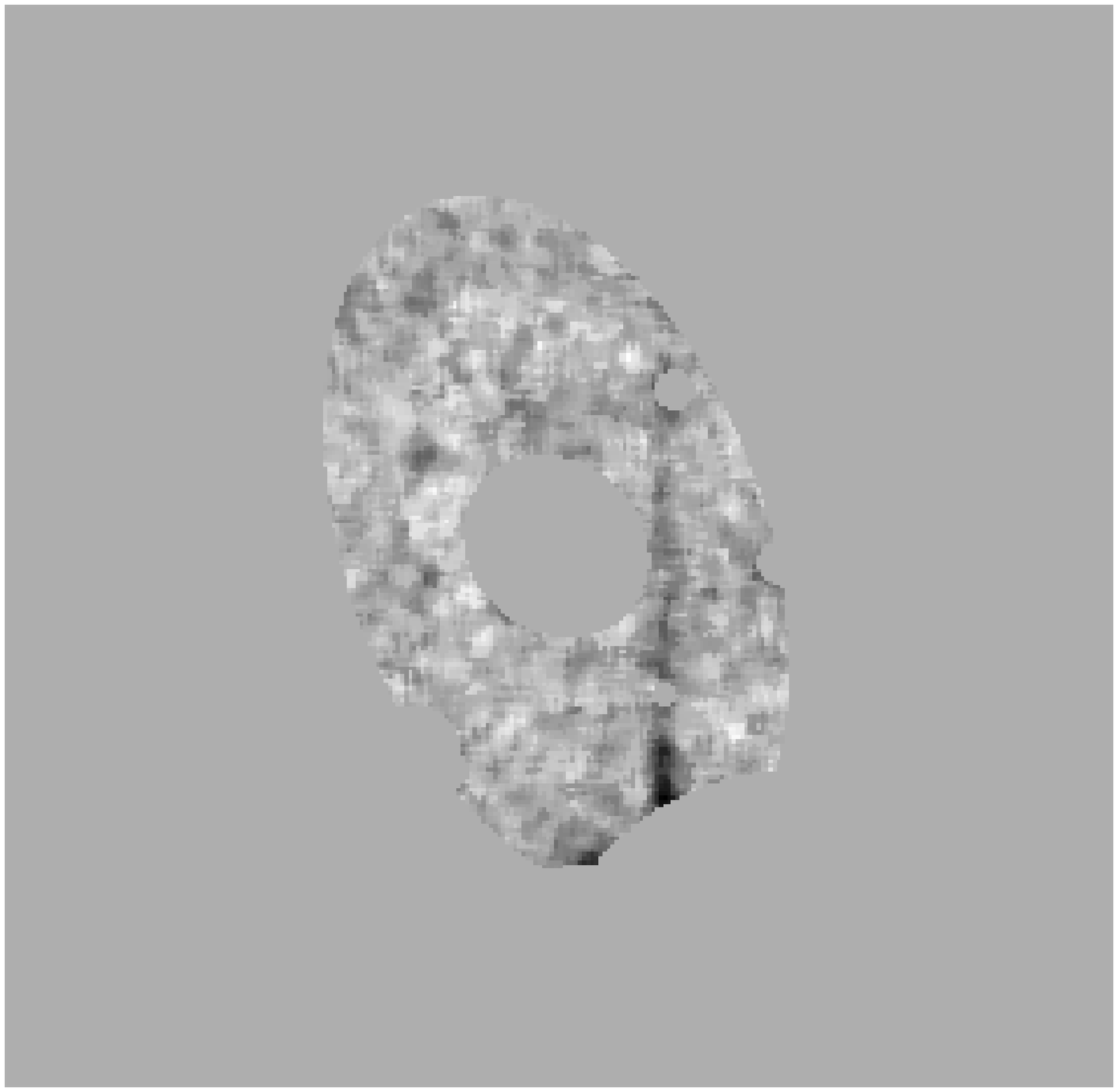}
  }
  \caption{Example of galaxy in Abell 795 discarded interactively due to unmasked diffraction spike.  Left column: galaxy image displayed with 15" scale bar.  Right column: residual of galaxy image divided by model with galaxy core and neighboring external objects masked.  The tidal parameter used to select tidally-disturbed galaxies uses the mean value of the unmasked areas of this image together with a noise correction frame.\label{fig-spike}}
\end{figure}

\begin{figure}
  \centerline{
    \includegraphics[width=2.67cm, angle=0]{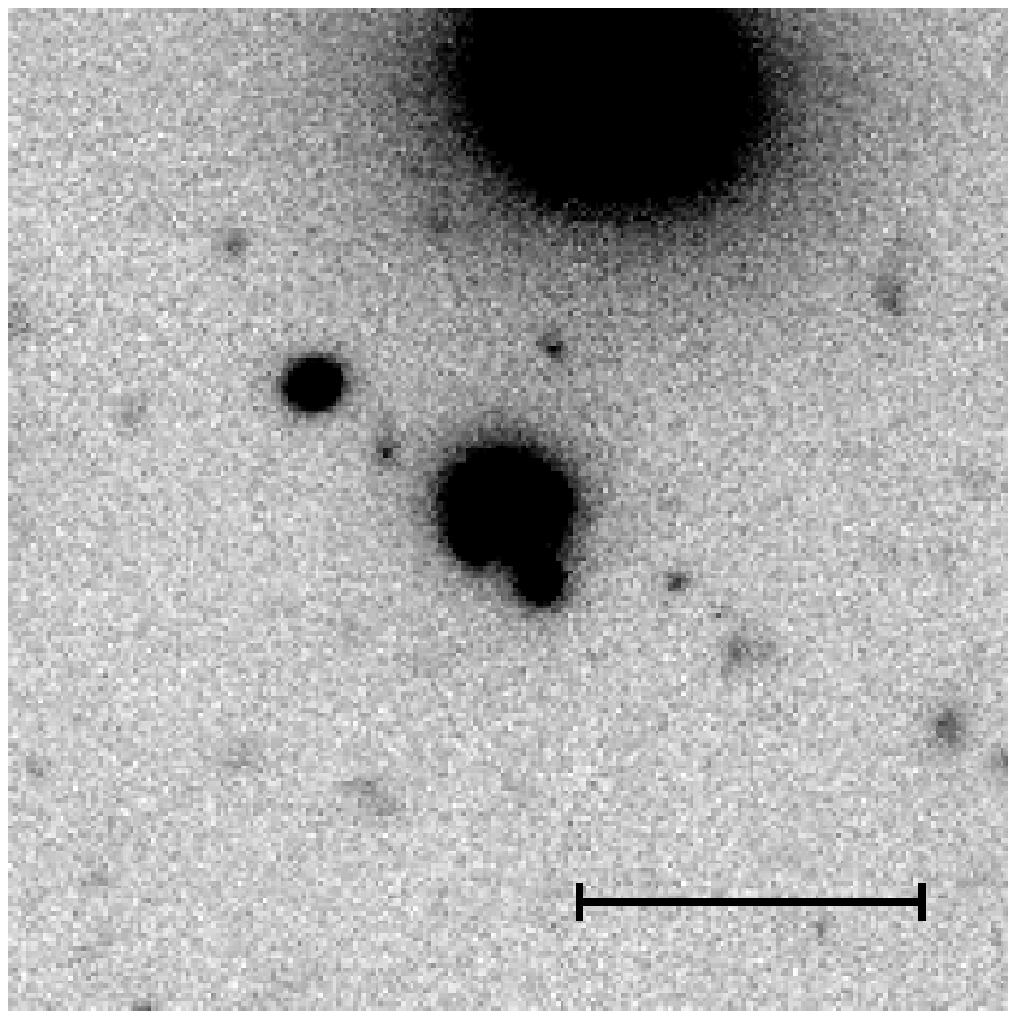}%
    \includegraphics[width=2.67cm, angle=0]{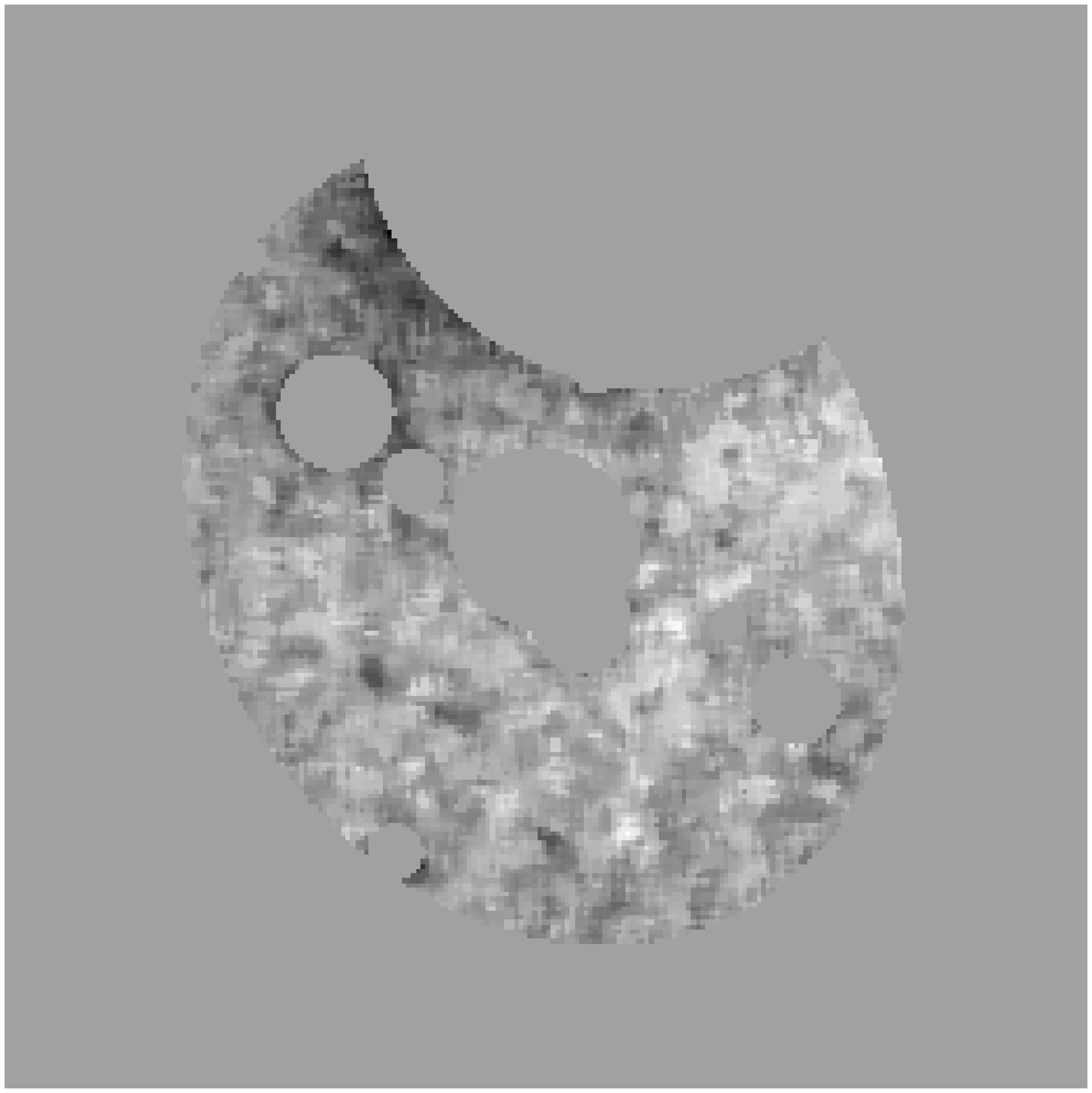}%
    \includegraphics[width=2.67cm, angle=0]{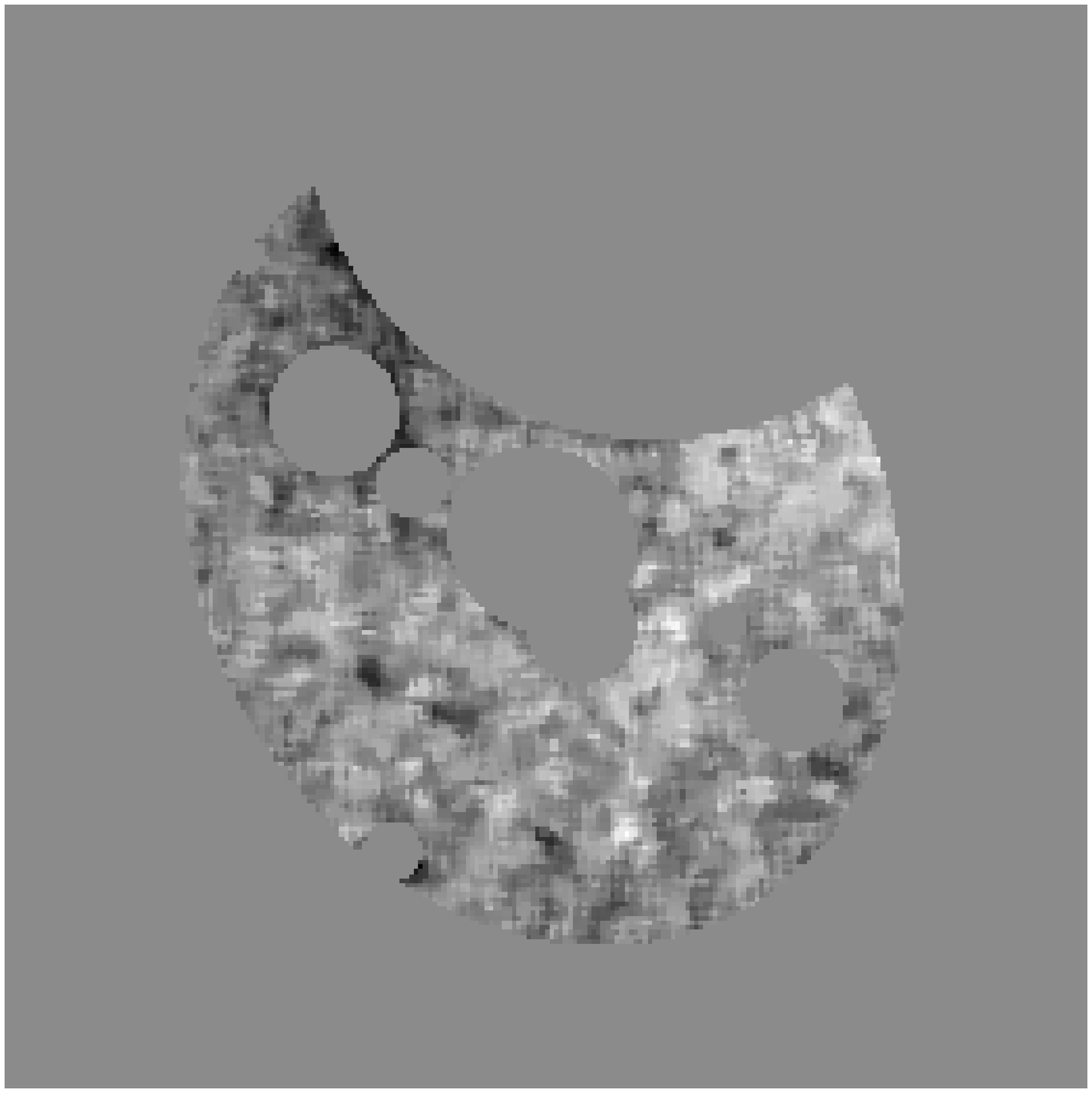}
  }
\caption{Sample galaxy in Abell 795 automatically discarded due to extended light profile of bright nearby galaxy.  Galaxies whose tidal parameters decrease by more than 5\% when the mask sizes for bright sources are increased by 10\% are discarded as this is an indication that the high tidal parameter is likely due to extended light from neighboring bright sources rather than from tidal features.
Left panel: Galaxy image shown with 15" scale bar, center: residual image, right: residual with larger masking.  Percent change in tidal parameter with larger masking: $-$16\%.\label{fig-icl}}
\end{figure}

The calculation of a tidal parameter does not by itself identify a set of
tidally disturbed galaxies because there is no absolute reference. We define a
threshold $T_c$ parameter that is best matched to what we visually identify as
tidally disturbed galaxies. The correspondence is not perfect, as we describe
further below, but we settle on defining those galaxies with $T_{c} > 9 \times
10^{-4}$ as tidally-disturbed.  We find that many galaxies have higher values
for the tidal parameter because light from the extended haloes of nearby bright
galaxies extends beyond the mask.  In an effort to eliminate these
false-positives we reject from our analysis candidate tidally-disturbed galaxies
whose tidal parameters decrease by more than 5\% when the mask sizes for bright
sources are increased by 10\% (e.g. Fig. \ref{fig-icl}).  While it is true that
this filter is biased against galaxies in dense regions, it would only bias our
results if real tidal features are being masked by the increased mask sizes.  We
verified that the fraction of discarded galaxies that were visually-disturbed
was consistent (within statistical errors) with the fraction of all galaxies
that are visually-disturbed within a given radial bin, indicating that these
filters do not introduce a radial trend.  The remaining galaxies with $T_{c} > 9
\times 10^{-4}$ make up the auto-detected sample of tidally-disturbed galaxies.
Lastly, because some false positives arise due to image quality issues, unmasked
sources, and diffuse light gradients from nearby bright galaxies, we visually
inspect this sample to construct our {\sl cleaned}-auto-detected sample of
tidally-disturbed galaxies.    The procedure we have just described consists of
many steps, all of which affect the number of systems that will ultimately be
classified as mergers, and thereby complicate comparison to models on an
absolute scale.  However, these filters should be mostly insensitive to
clustercentric radius, particularly outside the crowded central part of a
cluster right near the brightest cluster galaxy, and as such are not expected to
affect our results. 

\begin{figure}
  \centerline{
    \includegraphics[width=4cm, angle=0]{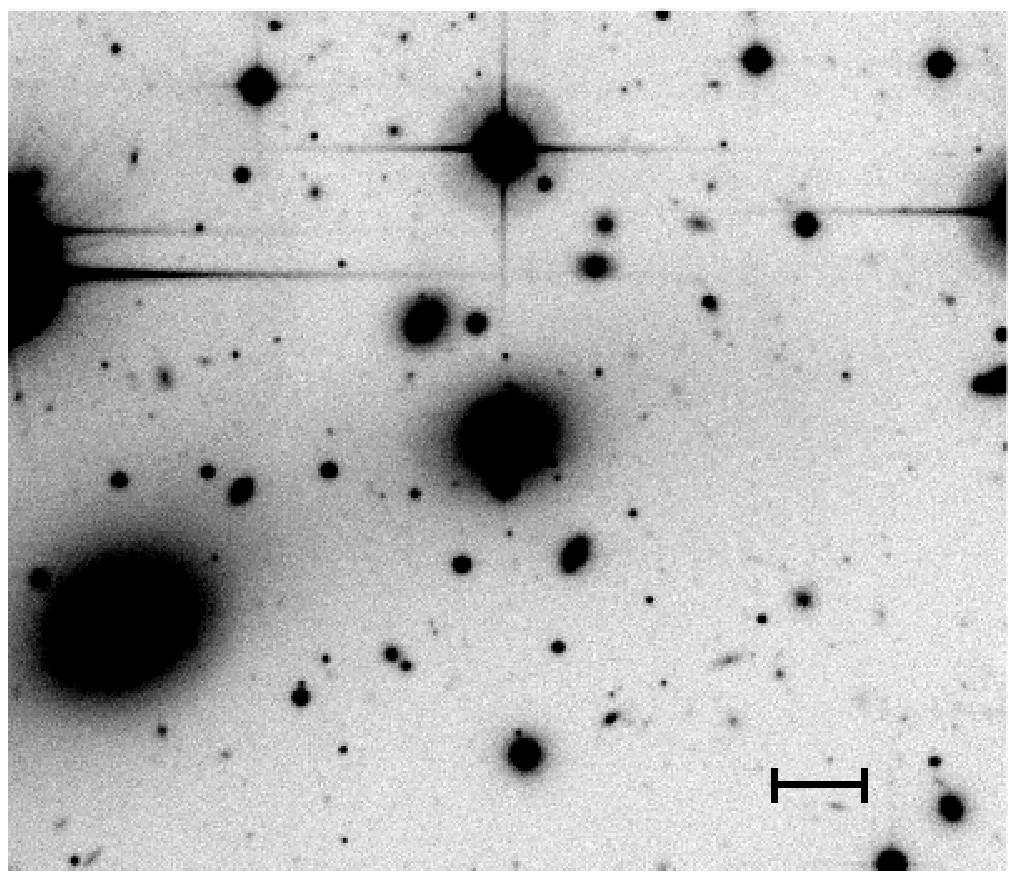}%
    \includegraphics[width=4cm, angle=0]{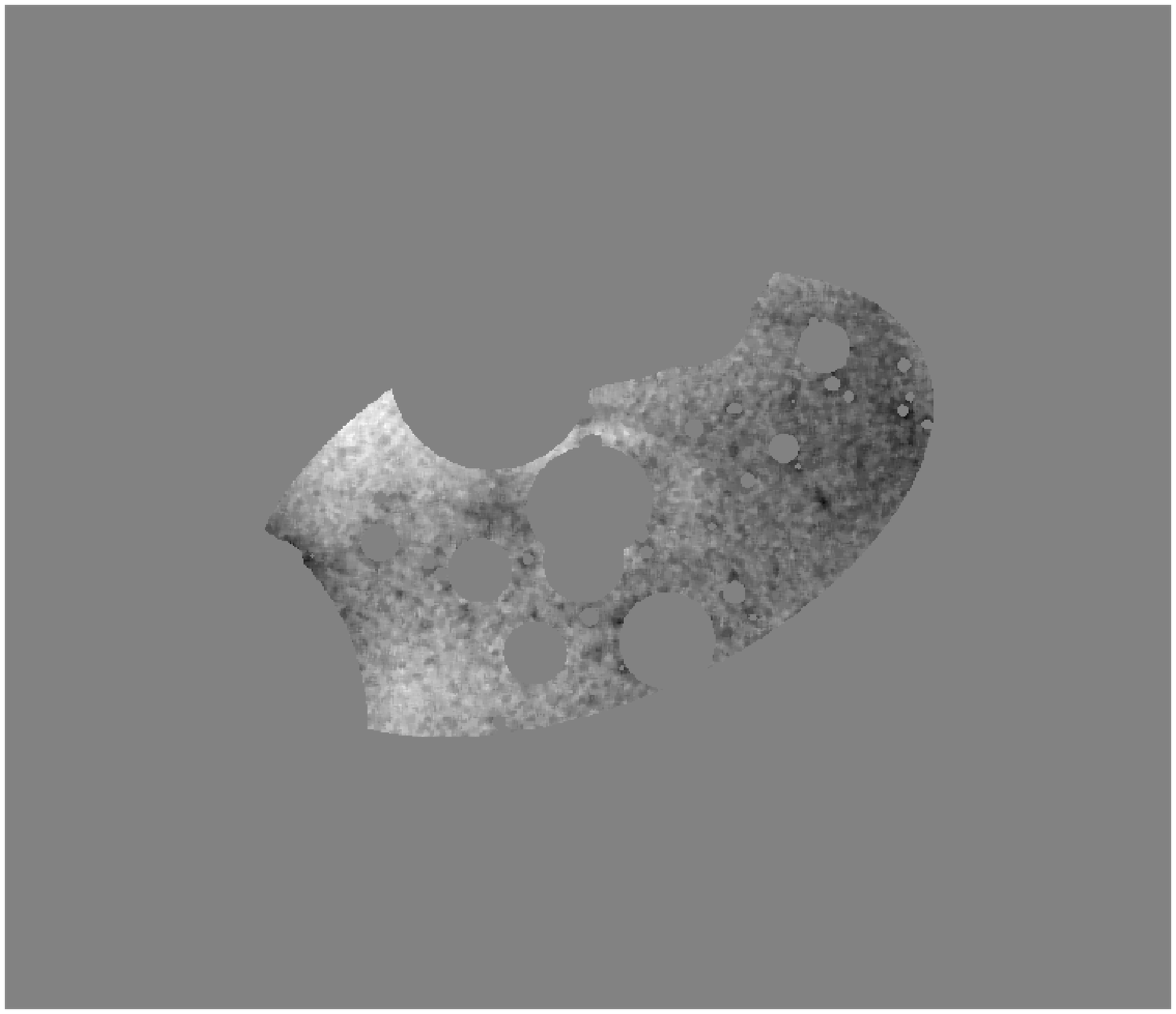}
  }
  \centerline{
    \includegraphics[width=4cm, angle=0]{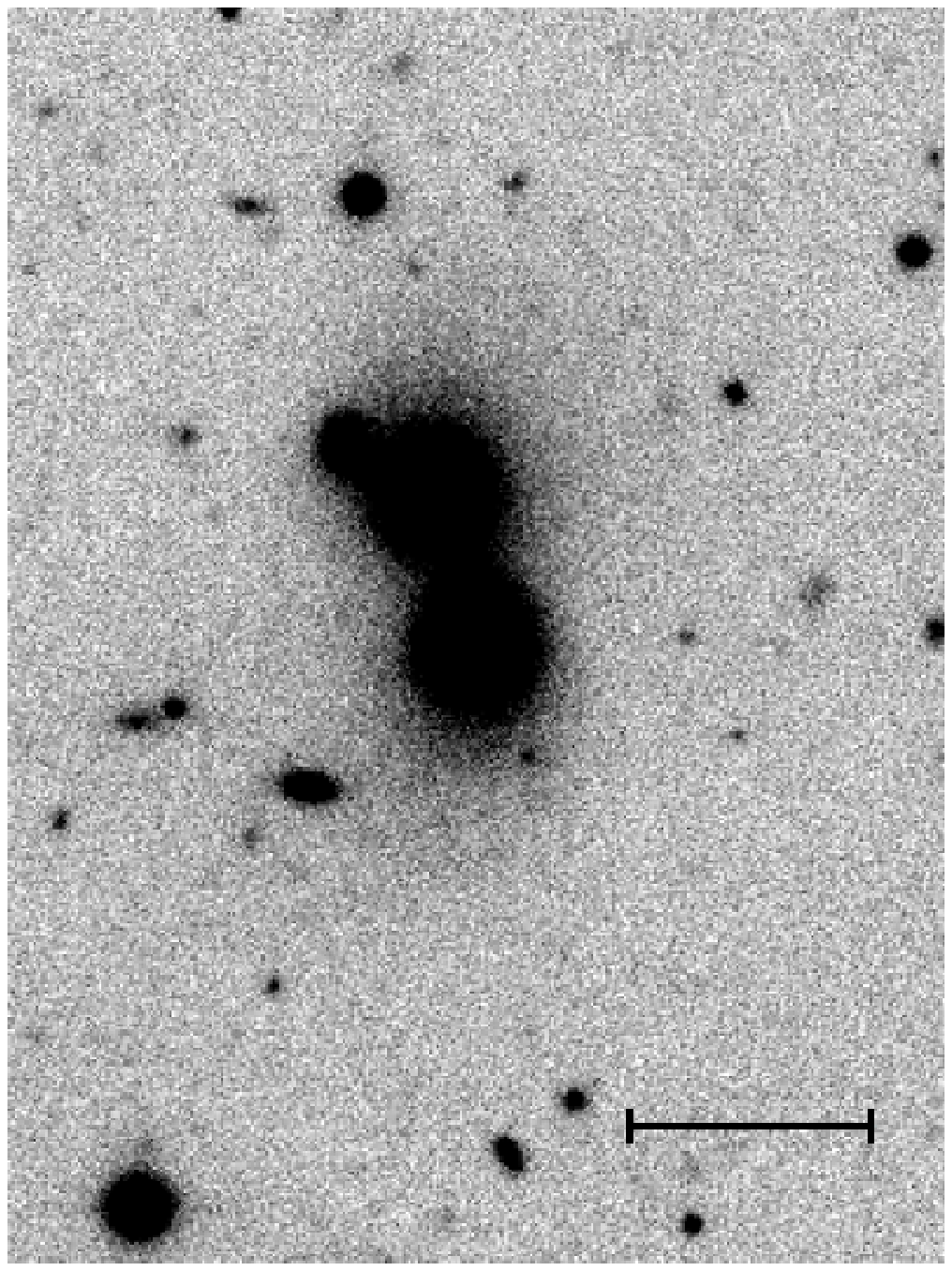}%
    \includegraphics[width=4cm, angle=0]{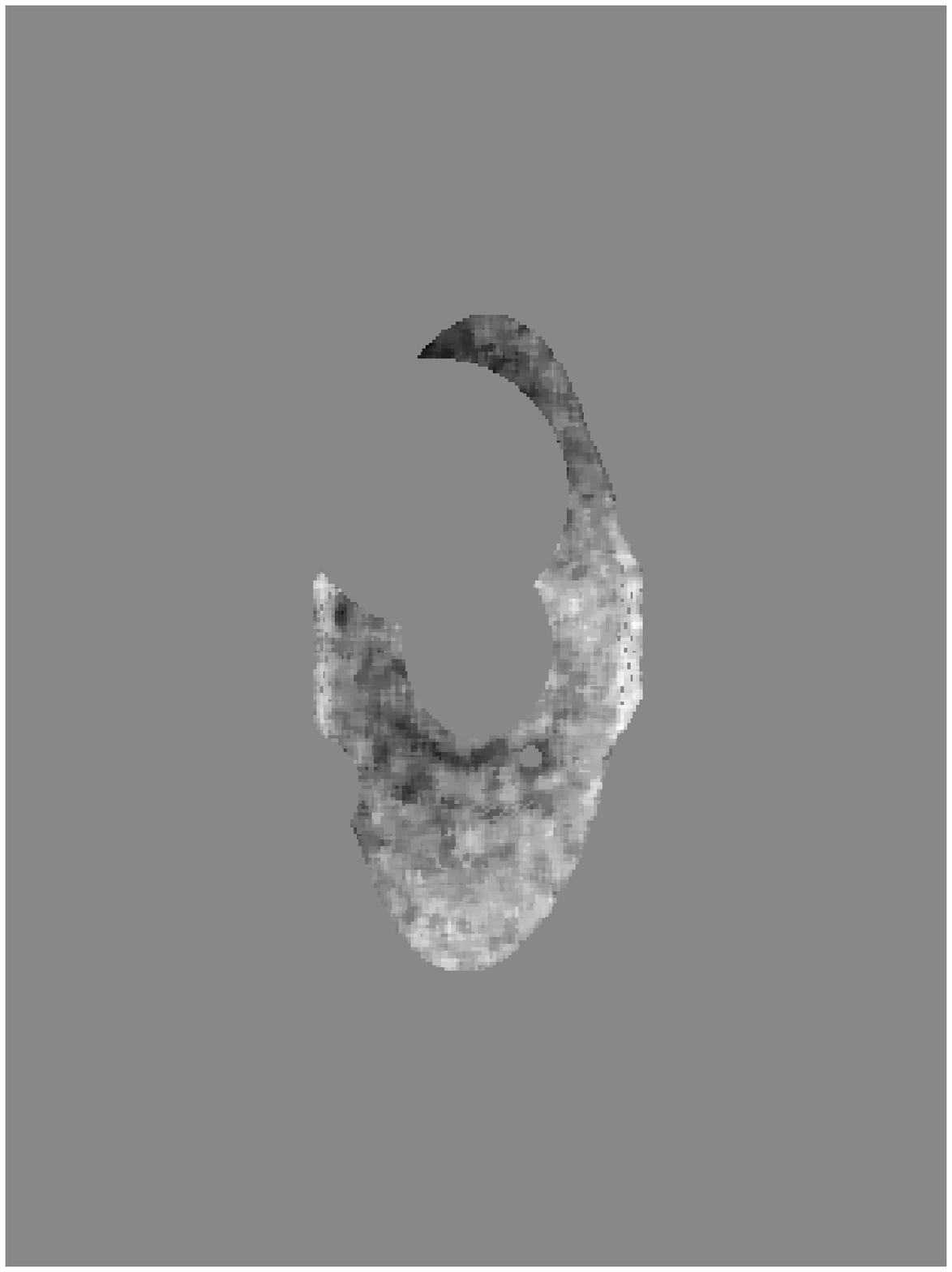}
  }
  \caption{Sample galaxies flagged as disturbed in the auto-sample, but excluded from the cleaned-auto sample in Abell 553 (top) and Abell 401 (bottom).  Left column: galaxy images displayed with 15" scale bars, right: region used for the residual calculation that has the ellipticity of the primary galaxy and has regions masked.  Top panel: it is not clear whether these two galaxies are interacting or are a close superposition.  The high mean residual value appears to be primarily caused by light from the adjacent galaxy spilling beyond its mask rather than any visible tidal structure ($T= 1.4 \times 10^{-3}$).
Bottom panel: the high residual value appears to be primarily due to the extended light profile of a nearby giant elliptical galaxy ($T = 1.8 \times 10^{-3}$).\label{fig-auto_fail}}
\end{figure}

\begin{figure}
  \centerline{
    \includegraphics[width=4cm, angle=0]{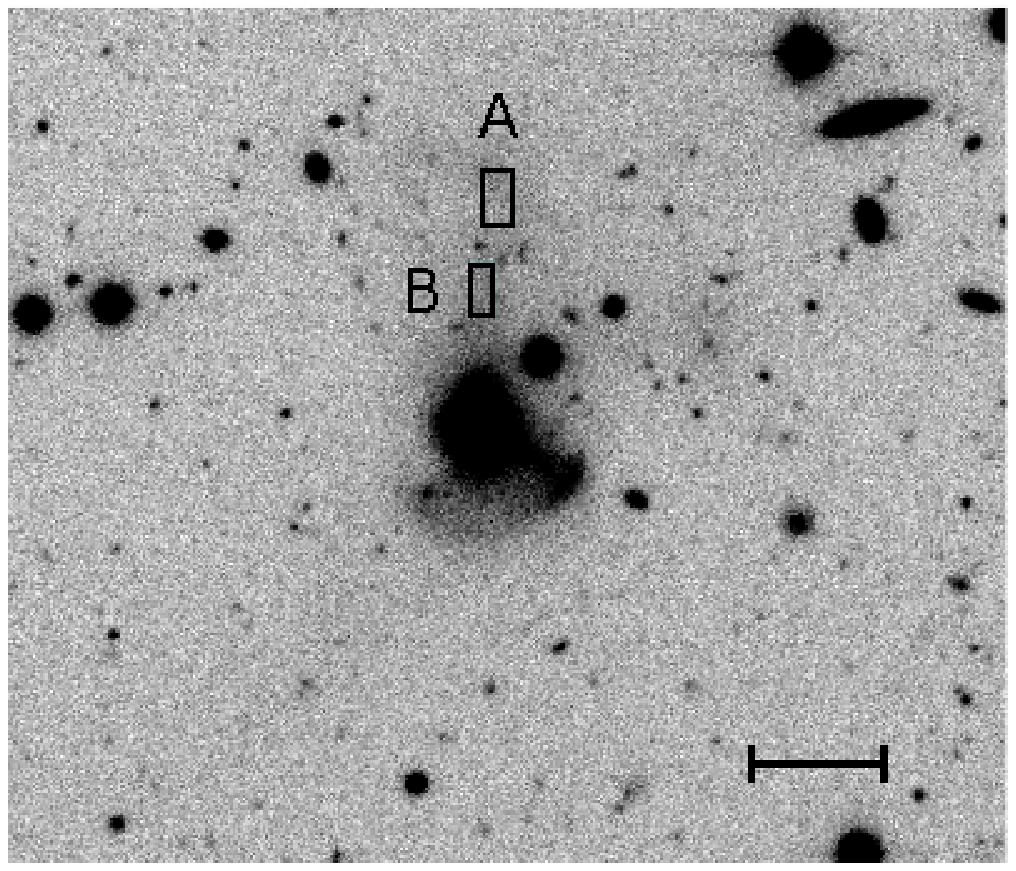}%
    \includegraphics[width=4cm, angle=0]{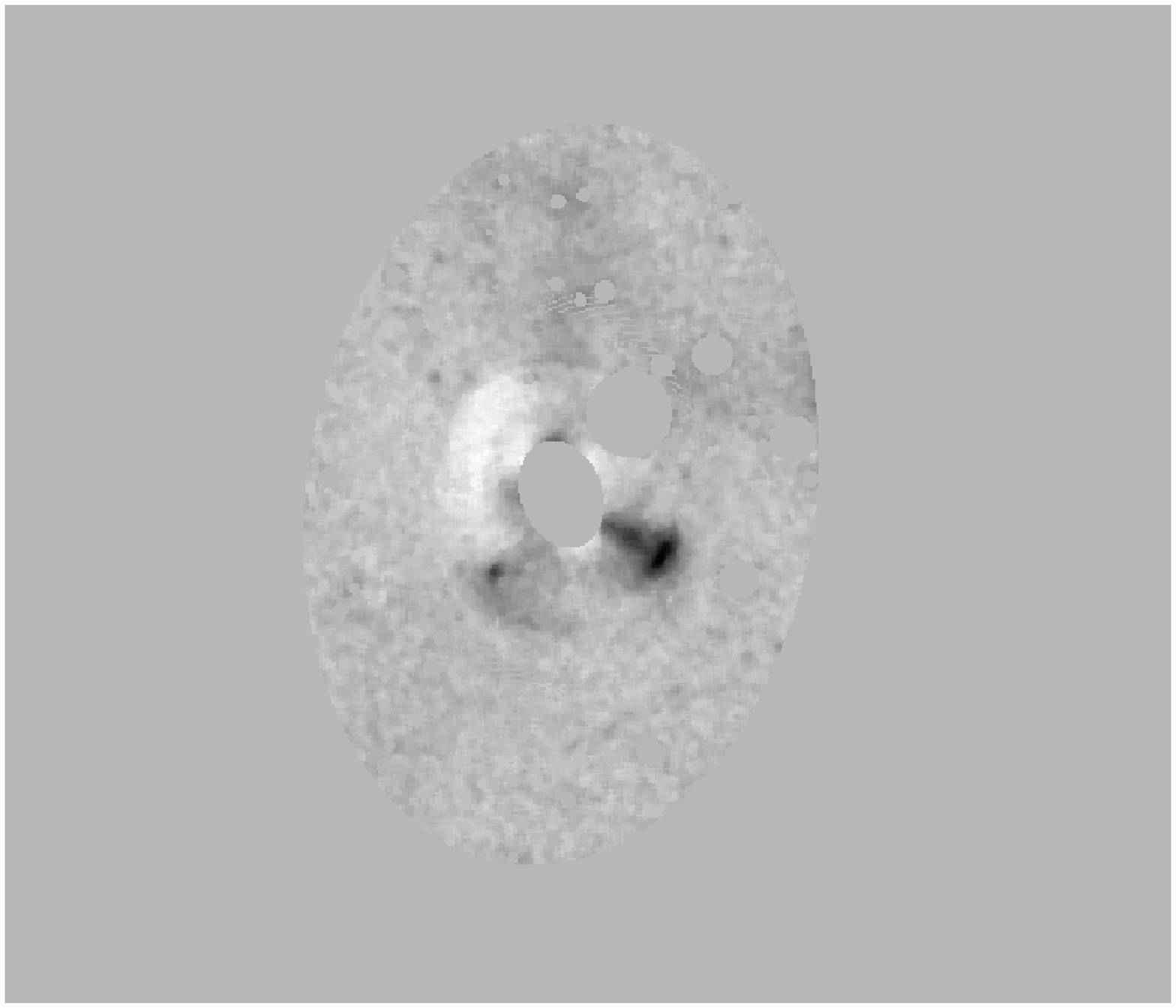}
  }
  \centerline{
    \includegraphics[width=4cm, angle=0]{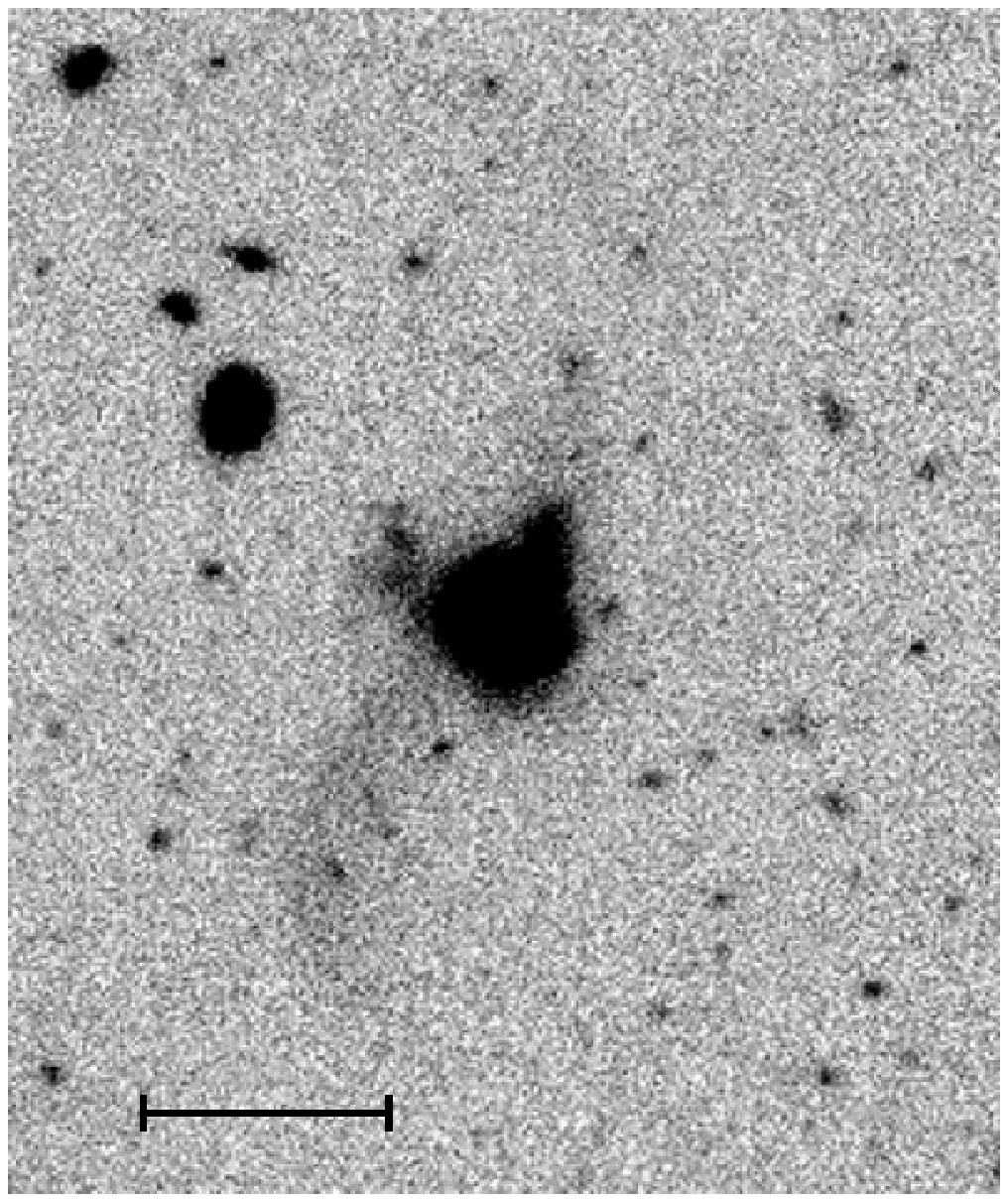}%
    \includegraphics[width=4cm, angle=0]{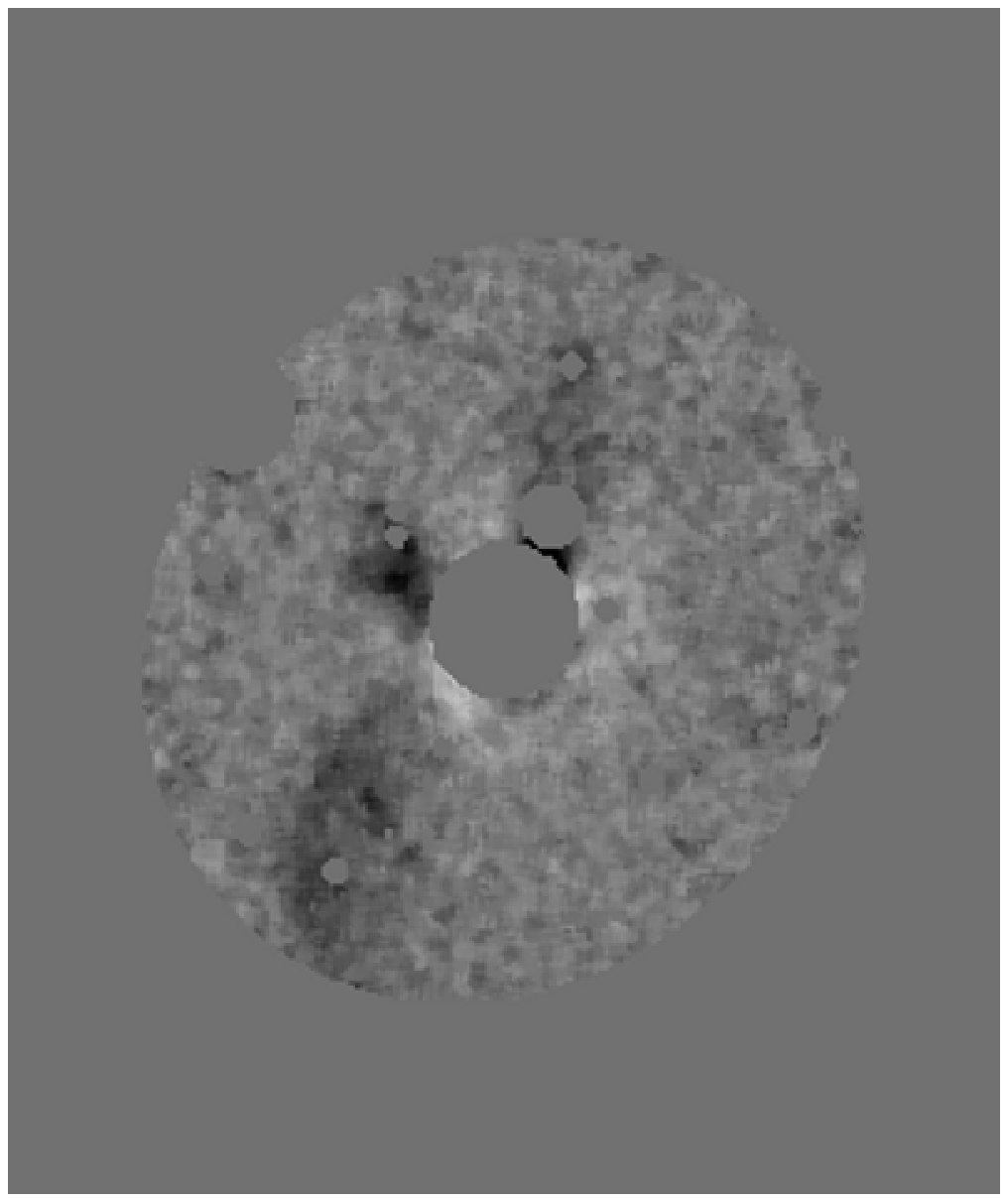}
  }
  \caption{Examples of galaxies marked as disturbed in both the visual and cleaned-auto samples in Abell 21 (top) and Abell 1068 (bottom) .  Left column: galaxy images displayed with 15" scale bars, right: region used for the residual calculation that has the ellipticity of the primary galaxy and has regions masked.  Top panel: the galaxy is clearly undergoing a merger; note the bright shell below the galaxy and the diffuse tidal stream above the galaxy ($T = 1.2 \times 10^{-3}$).  The mean surface brightness within the red box ``A" is 25.4 and within box ``B" is 25.1.  Bottom panel: also a clear case of a recent or ongoing merger with a distinct tidal tail extending to the lower left of the galaxy and tidal debris to the left and above the galaxy ($T = 1.1 \times 10^{-3}$).  Galaxies that are identified by both the visual and cleaned-auto samples tend to have the strongest and most distinctive tidal features.\label{fig-visual_and_auto}}
\end{figure}

\begin{figure}
  \centerline{
    \includegraphics[width=4cm, angle=0]{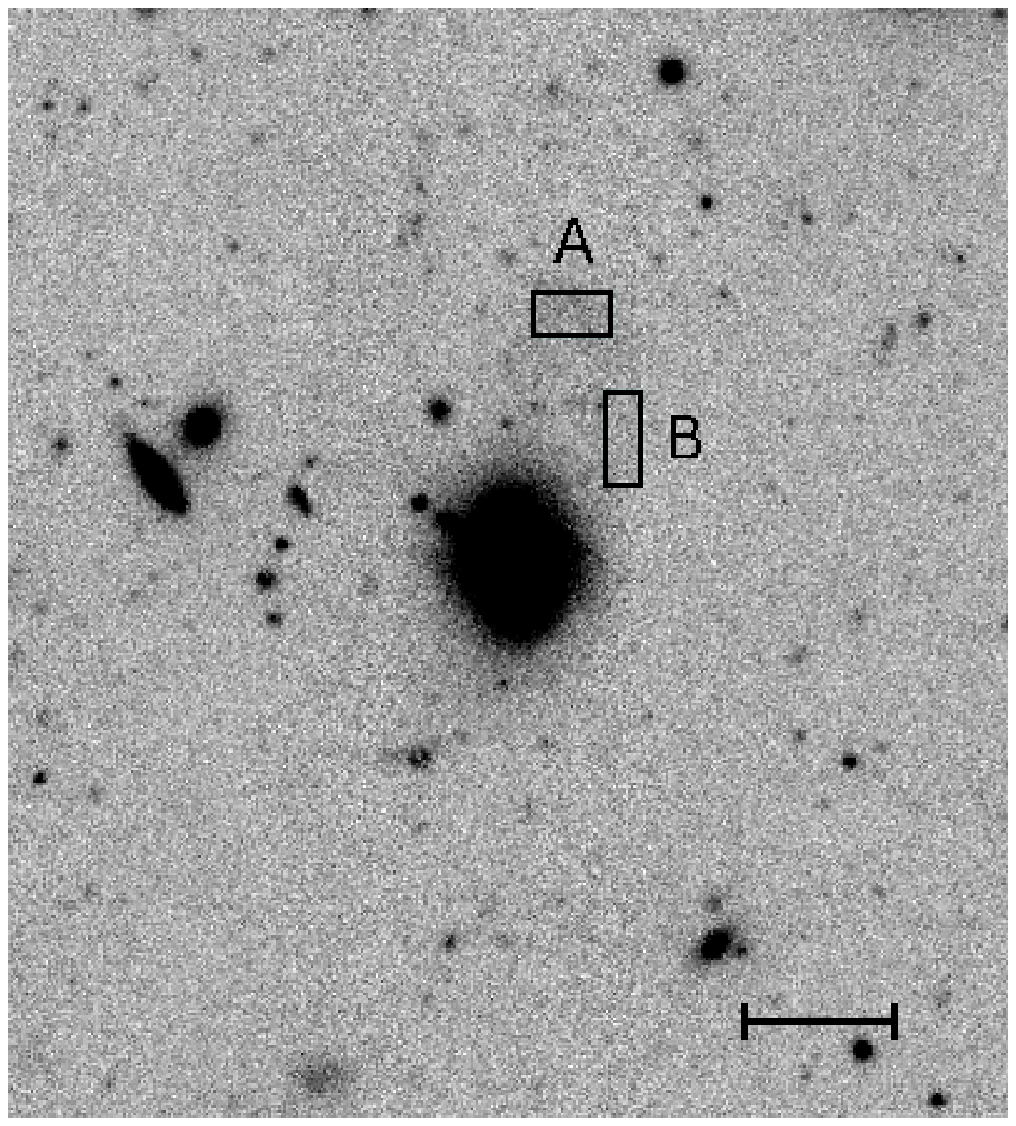}%
    \includegraphics[width=4cm, angle=0]{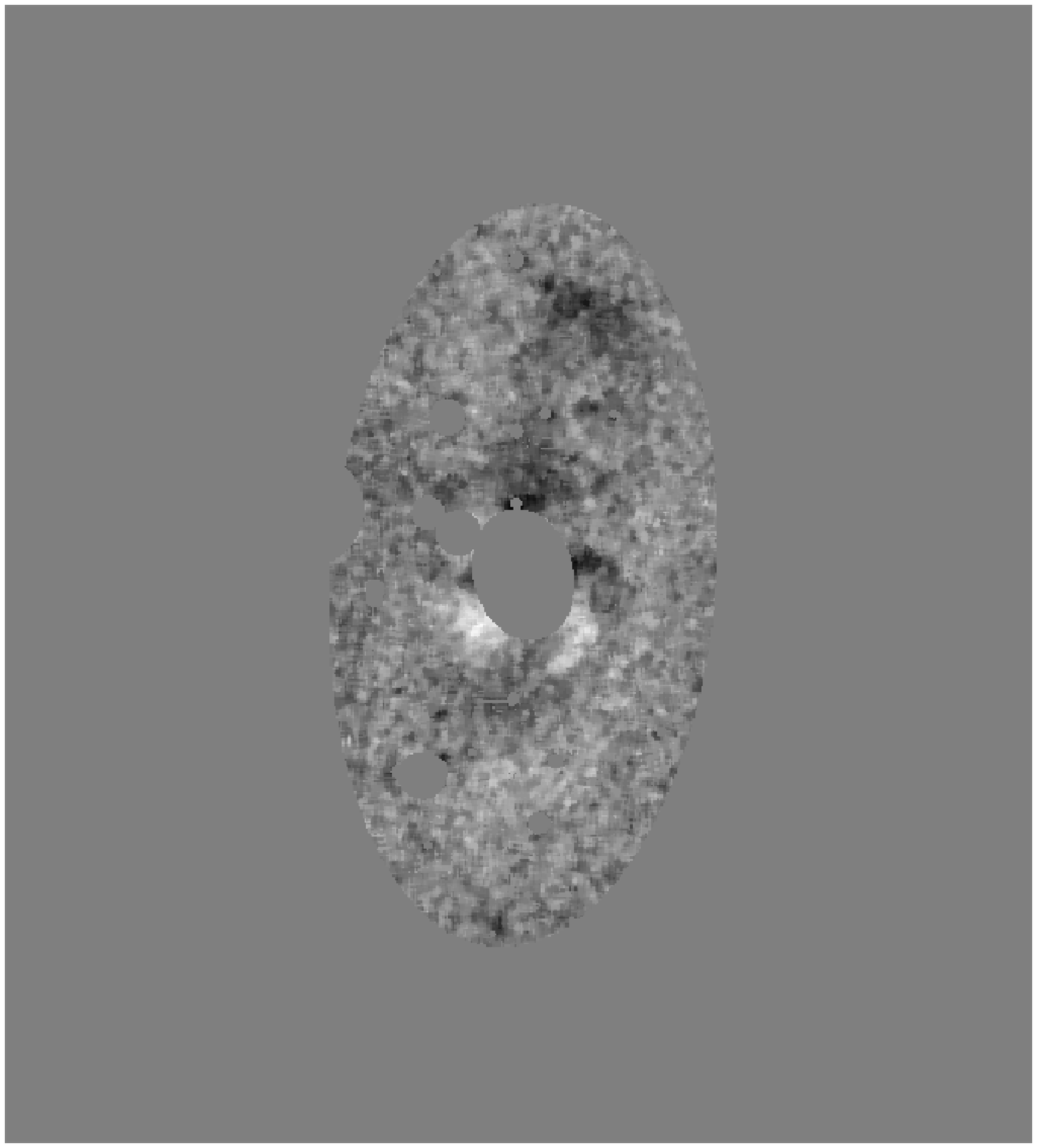}
  }
  \centerline{
    \includegraphics[width=4cm, angle=0]{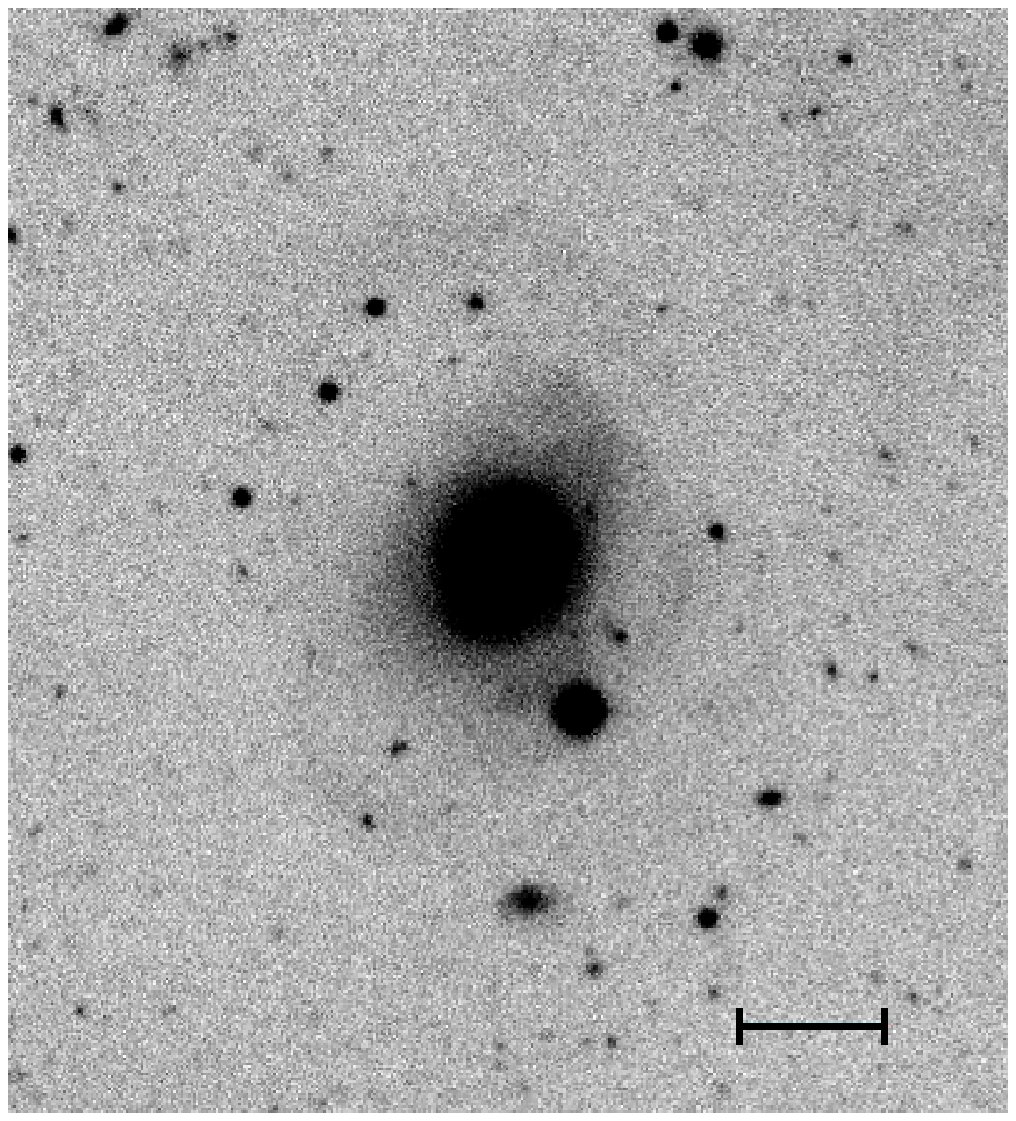}%
    \includegraphics[width=4cm, angle=0]{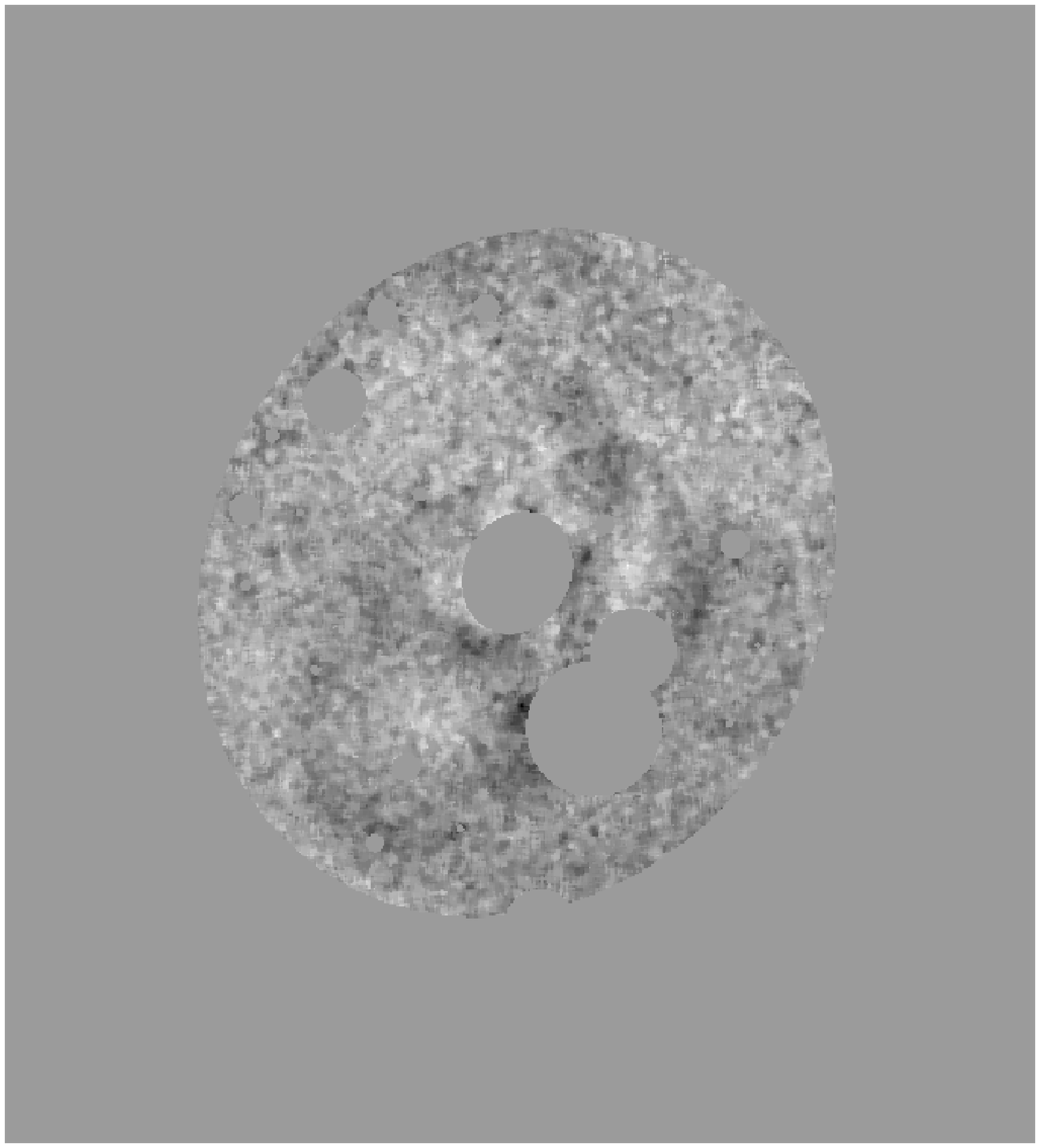}
  }
  \centerline{
    \includegraphics[width=4cm, angle=0]{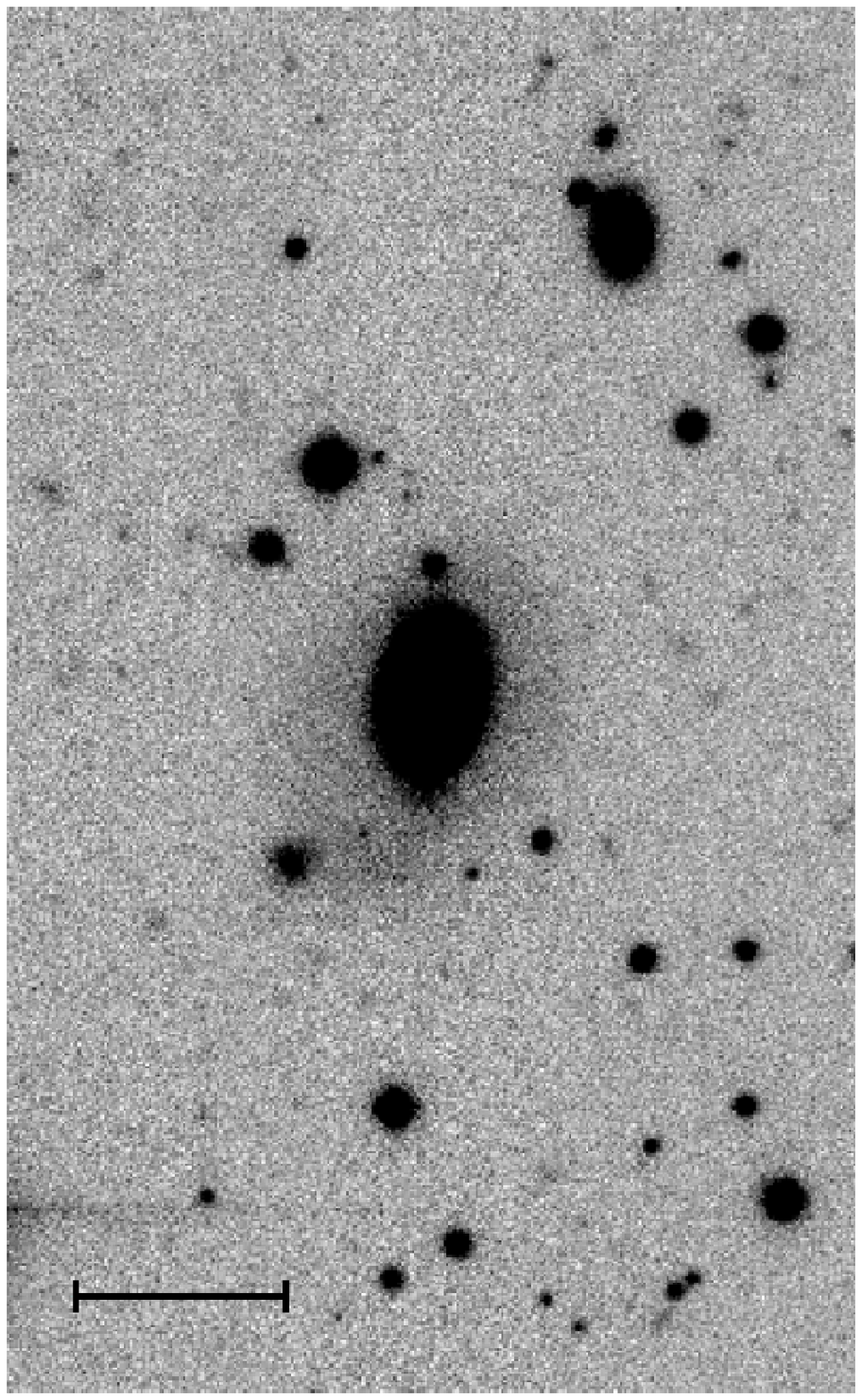}%
    \includegraphics[width=4cm, angle=0]{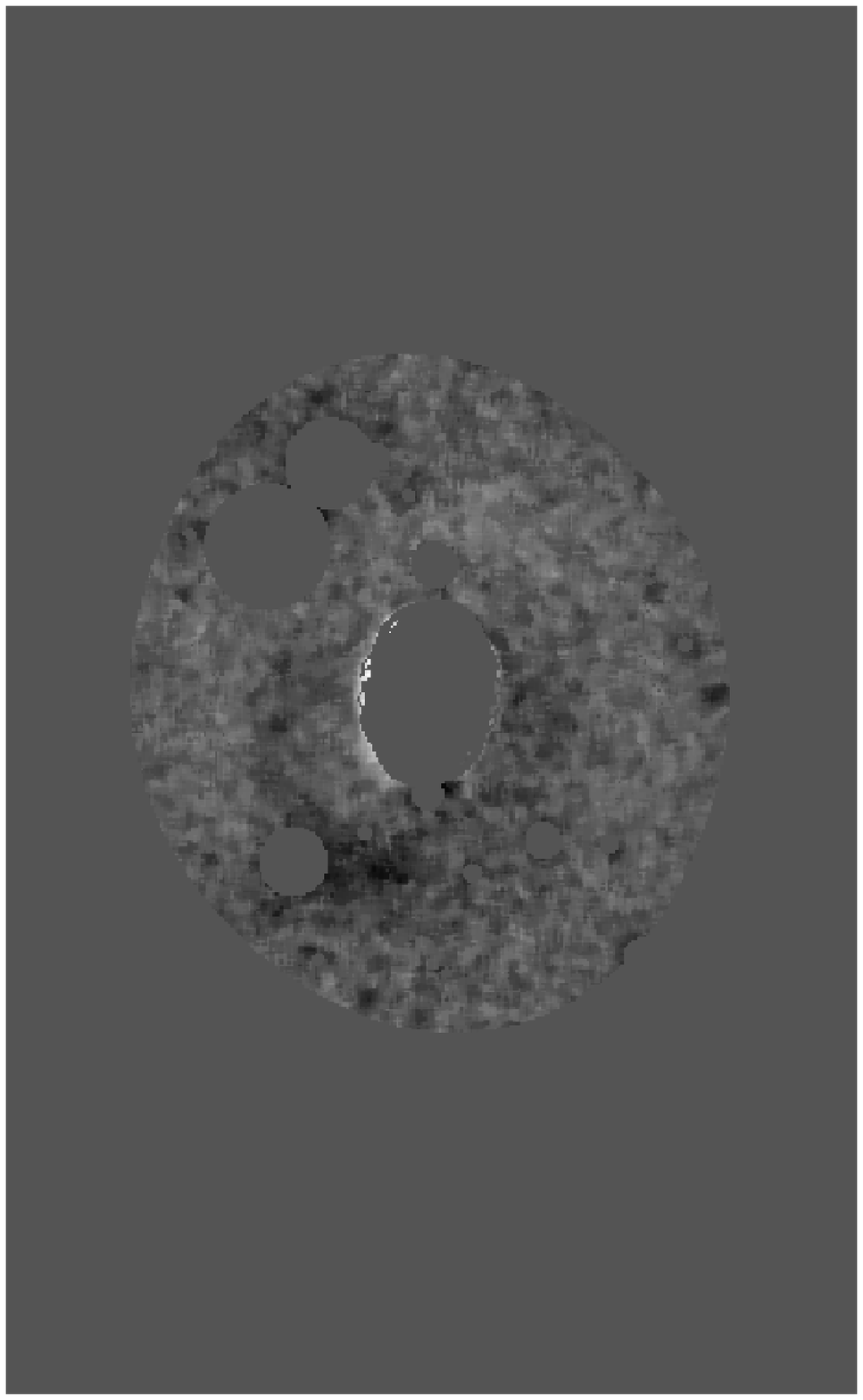}
  }
  \caption{Examples of galaxies in either the visual or cleaned-auto samples (but not both) in Abell 399 (top and middle panels) and Abell 553 (bottom).  Left column: galaxy images displayed with 15" scale bars, right: region used for the residual calculation that has the ellipticity of the primary galaxy and has regions masked.  Top panel: Sample galaxy flagged as disturbed in the visual sample, but not in the cleaned-auto sample ($T = 8.6 \times 10^{-4}$).  Note the ring of tidal debris extending from the top region of the galaxy.  The mean surface brightness within the red box ``A" is 25.7 and within box ``B" is 26.3.  Middle panel: Another sample galaxy flagged as disturbed in the visual sample, but not in the cleaned-auto sample ($T = 7.1 \times 10^{-4}$).  Note the series of tidal shells.  Bottom panel: Sample galaxy flagged as disturbed in the cleaned-auto sample, but not in the visual sample ($T = 9.1 \times 10^{-4}$).  Note the diffuse shell extending from the lower left portion of the galaxy.\label{fig-visual_or_auto}}
\end{figure}

3710 galaxies satisfy all of the automated selection criteria (cluster
color-magnitude band, $M_{r^\prime}< -20$, $r^{-1/4}$ radial brightness profile,
successful IRAF $bmodel$ generation, and less than 50\% of the area surrounding
the galaxy masked). Of these, we discard 79 that are problematic due to masking
issues and background light gradients (e.g. Fig. \ref{fig-spike}).  Of the
remaining 3631 galaxies, 202 were flagged as potentially-disturbed by our tidal
parameter threshold, while 109 galaxies were visually-identified as
tidally-disturbed (see Figures \ref{fig-visual_and_auto} and
\ref{fig-visual_or_auto} for examples).  While there are a large number of
disturbed galaxies that the automated sample identified that were missed by the
visual inspection (see Fig. \ref{fig-visual_or_auto} for an example), the
sample also includes many false identifications (usually due to image quality
problems or unmasked sources).  By eye we discard obvious false detections from
the automated sample to make a cleaned-auto sample of 122 galaxies (see Fig.
\ref{fig-auto_fail}) out of 3551 total galaxies (hereafter referred to as
``all"). The samples are compared below.

\section{Results}

\subsection{Rates of Tidal Features}

The intersection of the visual and cleaned-auto samples includes 42 galaxies,
while the union of the two samples includes 189 galaxies. There is no single
characteristic that uniquely determines whether a galaxy is identified or missed
by the visual or cleaned-auto samples.  The visual sample includes galaxies that
have tidal features that are close to ellipsoidal, of smaller spatial extent, or
of such low surface brightness that they are missing from the cleaned-auto
sample.  The cleaned-auto sample includes galaxies whose residual images show
strong features that were not included in the visual sample because it was not
evident whether these are actual tidal features,  spiral arms, or nearby
satellite galaxies. There are also a few instances of the cleaned-auto sample
missing galaxies whose features are very prominent, because these have been
masked as separate sources by SExtractor. The intersection sample tends to
include the strongest, most spectacular features, while the union is a more
complete sample, but suffers from a larger fraction of questionable detections.  

The tidal feature rate among early-type galaxies in these environments, using
the cleaned-auto and visual samples, is $3.4 \pm 0.3\%$ and $3.0 \pm 0.3\%$,
respectively, where the uncertainties are purely statistical. The union of the
two samples results in a rate of  $5.3 \pm 0.4\%$. As we    mention in \S 2.2 a
tight color selection could result in a bias if interacting galaxies are bluer
than the mean.  The mean color offset from the red sequence is $0.00 \pm 0.07$
for the visually-selected sample and $-0.01 \pm 0.06$ for   the cleaned-auto
sample, while the mean color offset for all galaxies analyzed is $0.00 \pm
0.06$, demonstrating that  our interacting samples do not suffer from a color
bias.  The rate of tidally disturbed cluster galaxies we find is significantly
lower than the rate observed by \cite{tal}, but the difference is likely because
our sample is $\sim$ an order of magnitude more distant, resulting in the
smaller apparent size of galaxies and their associated tidal features.  Our
rates match more closely those of \cite{bridge} and \cite{miskolczi}, although
again it is complicated to compare results among studies.

Although the measured rate of tidal features appears small at first, some minor
adjustments can lead to an inference that the majority of the initial cluster
elliptical population must have experienced a major merger. First, the $\sim$
3\% rate, implies that at least 6\% of the initial population experienced a
merger because a merger involves two galaxies. If one then considers that the
tidal signatures will not last indefinitely the numbers rise again. If, for
example, tidal features last 2 Gyr \citep{quinn} then the rate of mergers
increases by a factor of about 5 to $\sim 30\%$. If one further considers that
mergers were probably more likely in the past, then the number rises even more.
None of this is empirically well determined, but our fundamental conclusion is
that our low rate measurement does not necessarily imply a low incidence of
mergers over the lifetime of these galaxies.

\subsection{Interlopers and Projection Effects}

Before proceeding to examine the spatial distribution of galaxies with tidal
features, we stop to consider the effect of interloping galaxies on our radial
trends. Without spectroscopic redshifts, we do not know with certainty whether
or not a particular galaxy is in the cluster. This acts to muddle any real
correlations with radius.  To estimate the effect of interlopers within our
color-selected samples, we count the number of galaxies selected using
color-magnitude bands of the same size displaced redward by 0.26 mag (the
thickness of our selection area).  These displaced color-magnitude bands contain
2834 galaxies, while the bands centered on the red sequences contain 11,904
galaxies, suggesting that approximately 80\% of our selected galaxies are
cluster members.  We also count the number of galaxies in the four deep fields
of the CFHTLS \citep{gwyn} that would satisfy our galaxy selection criteria.
The count from the CFHTLS-Deep fields yielded a much lower estimation of the
number of interloping galaxies, but had large field-to-field variation.
Therefore, we conservatively use the interloper estimate from the displaced
color-magnitude bands and repeat our tidal parameter analysis for these redder
galaxies to determine what effect they may have on the results.  A higher
fraction of the background galaxies are ellipticals, because we are dealing with
brighter and redder galaxies, resulting in a slightly higher fraction of these
(34\% versus 30\% of the cluster sample) meeting all selection criteria and
contributing to the final results.  As expected for a background population, the
fraction of interlopers that are tidally disturbed does not correlate with
clustercentric radius --- demonstrating that crowding toward the center of the
cluster does not artificially create a radial effect.  However, interlopers are
a larger fraction of our ``cluster" sample at larger clustercentric radii (as
high as 45\% beyond $R_{200}$ vs less than 10\% within $R_{200}$) due to the
steep decrease in the number density of actual cluster galaxies at larger
clustercentric radii.  The background galaxies have slightly higher rates of
tidal features ($4.3^{+0.8}_{-0.7}\%$ visually-detected  and
$4.5^{+0.8}_{-0.7}\%$ auto-detected, with the statistical uncertainties
calculated using the Poisson single-sided  limits from \cite{gehrels}) than the
cluster sample ($3.1 \pm 0.3\%$ visually-detected and $3.3 \pm 0.3\%$ auto-
detected) suggesting that they might slightly influence any radial behavior.

\subsection{Clustercentric Radius}

To examine the rates of tidal features as a function of projected clustercentric
radius, we rescale the projected radii using the $R_{200}$ value of each cluster
from \cite{sand2}. These radii were estimated using the conversion between $L_X$
and $M_{200}$ found by \cite{reiprich} and the correspondence between $M_{200}$
and $R_{200}$.  Clustercentric radii are measured from the brightest cluster
galaxy (though we verified that the results do not change significantly when the
radii are measured from the X-ray center). 

\begin{figure*}[]
\begin{center}
\plottwo{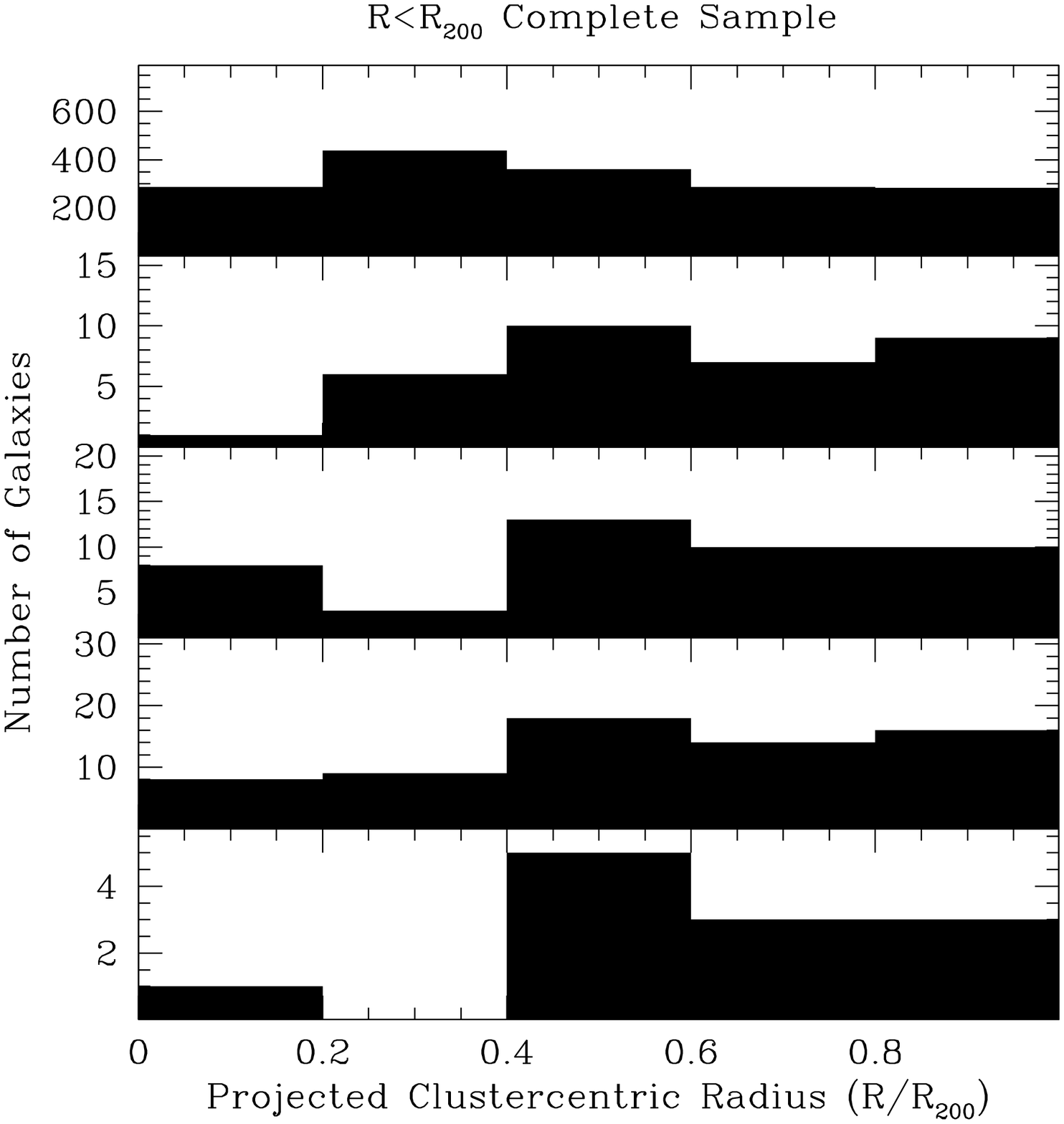}{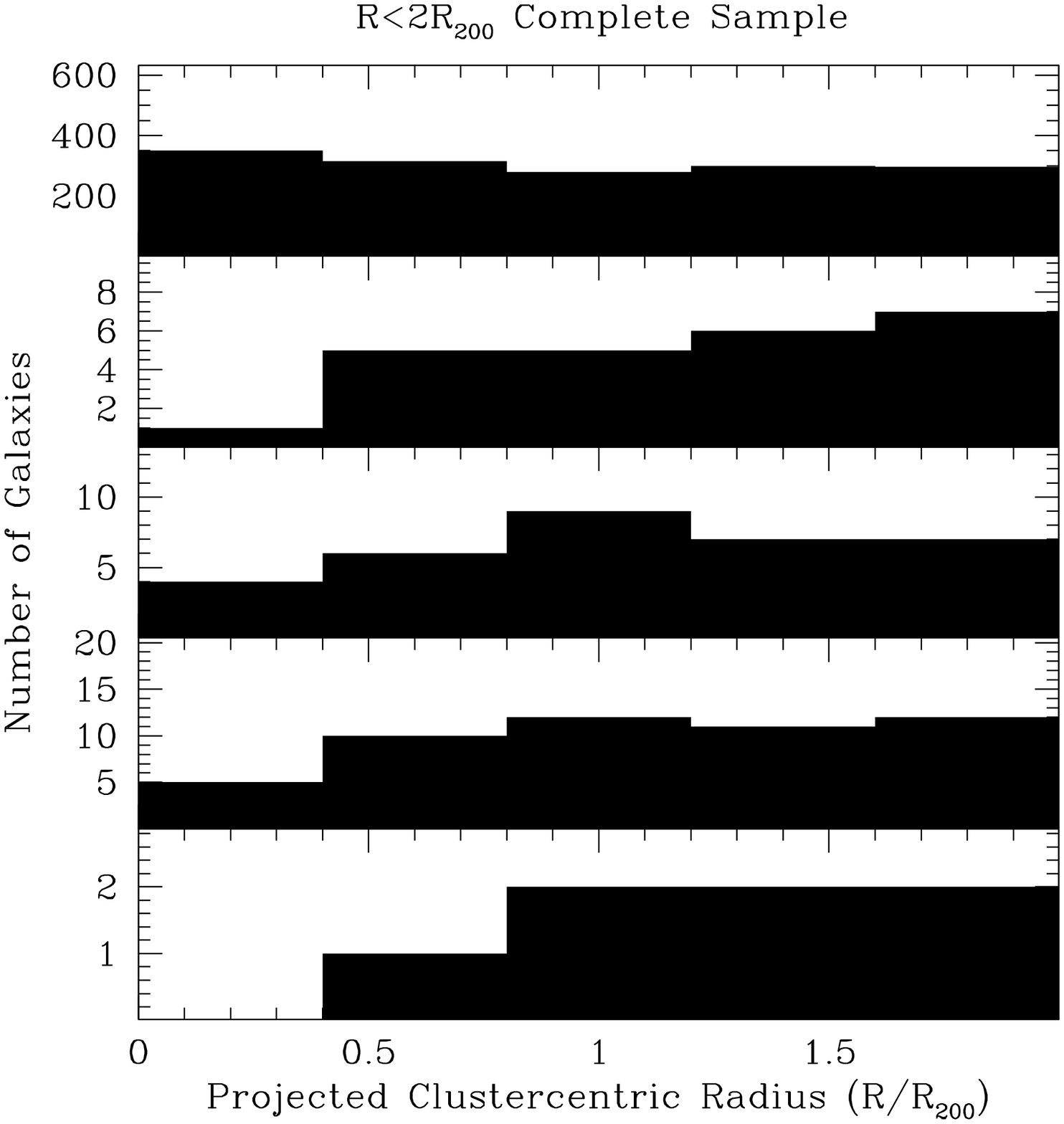}
\caption{Histograms of galaxies versus projected clustercentric radius.  The left section is the sample of 50 clusters for which the survey is complete within $R_{200}$, whereas the right section is the sample of 26 clusters complete within $2R_{200}$.  Top panel: ``all" galaxies, 2nd panel down: visually-selected disturbed galaxies, 3rd panel down: cleaned-auto sample of disturbed galaxies, 4th panel down: union of visually and cleaned-auto-selected galaxies, bottom panel: intersection of visually and cleaned-auto-selected galaxies.  \label{fig-histograms}}
\end{center}
\end{figure*}

Not all clusters are sampled to the same normalized projected clustercentric
radius, so we consider different subsamples.  We create a first subsample of the
1657 galaxies within $R_{200}$ in the 50 clusters for which our survey is
complete  out to $R_{200}$, a second subsample of the 1540 galaxies within
$2R_{200}$ in the 26 clusters for which we are complete to $2R_{200}$, and a
combined sample of all galaxies within $R_{200}$ in clusters complete to
$R_{200}$ and all galaxies within $2R_{200}$ in clusters complete to $2R_{200}$.
In Figure \ref{fig-histograms} we plot the relative radial distributions of
``all" galaxies and our various samples of galaxies with tidal signatures.

The distributions shown in Figure \ref{fig-histograms} make clear that there is
a significant deficit of galaxies with tidal signatures towards the centers of
these clusters.  The deficit is evident within $\sim 0.5R_{200}$, which is well
outside an area about the brightest central galaxy where one might expect that
confusion could lead to lower detection rates.  In fact, as we find with a
control sample of background galaxies that shows no gradient in tidal features,
we are not affected by confusion in the center of these clusters (\S 3.2).
Finally, these findings are independent of whether visual or automated
techniques are utilized. 

To place these conclusions on a more solid statistical footing, we apply K-S
tests to evaluate the likelihood that the cumulative distributions of galaxies
with tidal features are drawn from the parent distribution of  ``all" galaxies.
For the sample complete within $R_{200}$ we find that the tidal sample was not
drawn from the parent sample with 98.2\% confidence for the visually-selected
sample, 91.5\% confidence for the automated selection, 94.5\% for the
intersection of the two samples, and 97.4\% for the union. For the sample
complete to $2R_{200}$, the confidence limits fall below $2\sigma$ significance.

The radial variations in the incidence of tidal signatures suggests that either
(or both) the creation or destruction of tidal features depends on
clustercentric radius.  More appropriately, the variation depends on a physical
characteristic that drives the phenomenon and which is itself correlated with
clustercentric radius. A straightforward hypothesis is that tidal signatures are
more easily erased in a dense environment due to shorter dynamical times.  While
this is likely to be true near the center of the clusters, where the
intracluster light is built up, by the time one is considering galaxies at
$R_{200}$ and beyond, it may seem unlikely that the global potential will strip
tidal material.

\begin{figure}[]
\begin{center}
\plotone{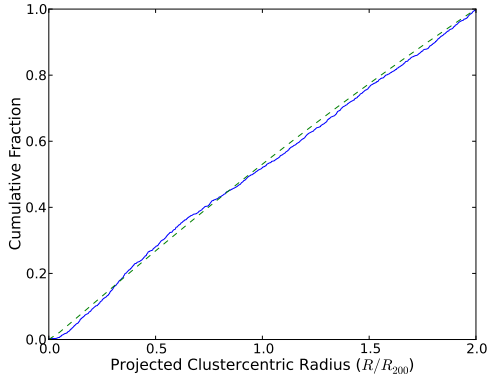}
\caption{Cumulative fraction of observed galaxies (within clusters with FOV complete to $2R_{200}$) as a function of projected clustercentric radius (solid blue line) compared to cumulative distribution of galaxies modeled by an NFW profile with c=2.6 and integrated out to a maximum radius of 10$R_{200}$ (dashed green line). \label{fig-distro}}
\end{center}
\end{figure}

To test this hypothesis, we rely on numerical simulations by \cite{rud09} that
find that the decay time of tidal streams is approximately 1.5 times the
dynamical time at their location in the cluster.  The dynamical time at
clustercentric radius $r$ is given by, \begin{equation} t_{dyn} =
\frac{\pi}{2}\sqrt{\frac{r^{3}}{GM}} \end{equation} where $M$ is the mass
enclosed within $r$.  We therefore set the lifetime of tidal features, $l(r)$,
to be $1.5 t_{dyn}$.  We determine the enclosed mass using the NFW density
profile \citep{nfw}: \begin{equation} \rho(r) =
\frac{\rho_{0}}{(r/a)(1+r/a)^{2}} \end{equation} where $ a = R_{200} / c $ and
we take $c = 2.6 $ to be an appropriate concentration parameter for the
distribution of galaxies in clusters \citep{bud}.  If the fraction of galaxies
to form tidal features, $f_{f}$, is uniform with clustercentric radius and time,
then the expected fraction of galaxies observed to have such features would be:
\begin{equation} f_{obs}(r) = \left\{ \begin{array}{ccl} f_{f} \frac{
l(r)}{t_{0}} & \mathrm{for} & l(r) < t_{0} \\ f_{f} &
\mathrm{for} & l(r) \geq t_{0}, \end{array}\right.  \end{equation}
where $t_0$ is the age of the Universe.  The observed fraction as a function of
projected clustercentric radius is then given by calculating the average
fraction along the line of sight: \begin{equation} f_{obs}(r_{proj}) =
\frac{\int ^{\infty} _{-\infty} f_{obs}(x) n(x) dx}{\int ^{\infty} _{-\infty}
n(x) dx} \label{model-eqn} \end{equation} where $ x = \sqrt{ r^{2}-r_{proj}^2} $
and $n$ is the number density of our ``all" sample.  Because the NFW profile is
not a good description of the mass or galaxy number density to arbitrarily large
radius, we replace the integration limit with $\pm 10 R_{200}$.  Using this
prescription and assuming that galaxies trace mass, we find that the
distribution of galaxies predicted as a function of projected radius by the NFW
profile roughly reproduces our observed distribution (see Fig.
\ref{fig-distro}).

\begin{figure*}[]
\begin{center}
  \plotone{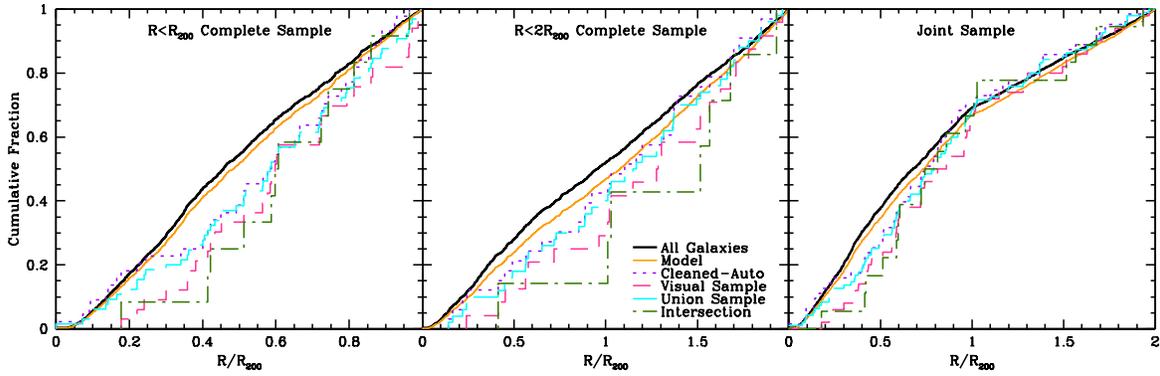}
  \caption{Cumulative fraction of galaxies versus projected clustercentric radius.  The left panel is the sample complete within $R_{200}$, the center panel is the sample complete within $2R_{200}$, and the right panel is the combination of the two samples (all galaxies within $R_{200}$ in clusters complete to $R_{200}$ and all galaxies within $2R_{200}$ in clusters complete to $2R_{200}$).  Solid black line: ``all" galaxies (dominated by undisturbed galaxies), dotted purple line: cleaned-auto sample, short-dashed red line: visually selected galaxies, long-dashed cyan line: union sample, dot-dashed green line: intersection sample, solid orange line: expected distribution for model (Eqn. \ref{model-eqn}) in which lifetime of the features is proportional to their dynamical time with a uniform incidence of features at all radii and an overall fraction of tidally disturbed galaxies set to match the observed value. All three panels show a deficit of tidally disturbed galaxies within $R_{200}$.\label{fig-cd_radius_model}}
\end{center}
\end{figure*}

We now compare our results to the expected distribution of observed tidal
features if the lifetime of the features is proportional to the local dynamical
time and the generation of tidal features is independent of time or
clustercentric radius (see Fig. \ref{fig-cd_radius_model}).  While the
cumulative distribution of tidally-disturbed galaxies still lies below the model
prediction within $R_{200}$, the deviation is still only statistically
significant for the visually-selected sample of disturbed galaxies.  For the
sample complete within $R_{200}$, the K-S test shows that the visually-selected
sample is drawn from a different distribution than the ``all" sample with 95.8\%
confidence, the automated selection differs with 81.3\% confidence, the
intersection of the two samples is discrepant at 91.0\% confidence, and the
union is discrepant at 91.7\%.  For the sample complete to $2R_{200}$ none of
the samples of disturbed galaxies are discrepant from the ``all" sample at the
$2\sigma$ level.  These results are robust with reasonable changes in the
concentration parameter used.  If we instead use c=6, which is more accurate for
the dark matter distribution for halos of this mass \citep{maccio}, the K-S
statistics are largely unchanged.  Additionally, we tested a range of
coefficients ($0.5-2.5 t_{dyn}$) for the decay time behavior of tidal streams
and, based on the one to two percent changes on the K-S probabilities, affirm
that our conclusions are independent of modest changes to the choice of the
coefficient. We find that this toy model, in which the lifetime of the features
is proportional to the local dynamical
time and the generation of tidal features is independent of time or
clustercentric radius, qualitatively agrees with the observed deficit of tidal features
at small clustercentric radii, but possibly would not solely account for the
magnitude of the deficit within $\sim 0.5R_{200}$.

\subsection{Local Density} When considering the effect of environment on cluster
galaxies there is always a tension between the role of global vs. local
environment. Part of the difficulty in addressing such issues is that the two
measures of environment are correlated, and therefore large samples are needed
to even begin to attempt to disentangle the two. A second difficulty is that
local environment is inherently difficult to measure in clusters due to the
complicated line-of-sight structure and relatively poor distance estimates. Here
we make an attempt to study the dependence of tidal features on local density,
but we face these same difficulties.

\begin{figure}[]
\begin{center}
\plotone{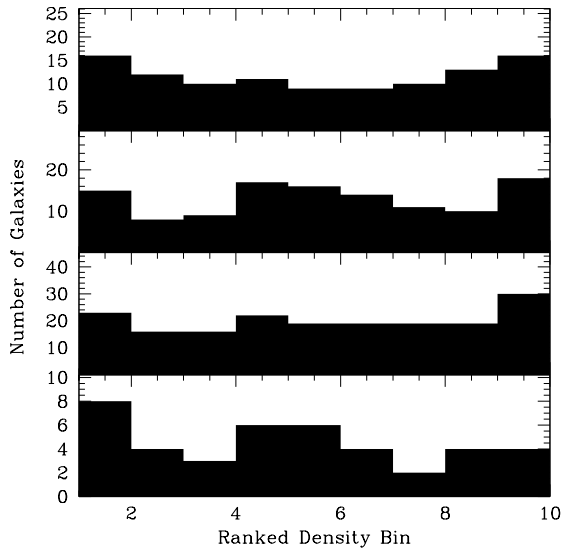}
\caption{Histograms of galaxies versus ranked local density in radius annulus (a random distribution would be flat across all density bins).  Top panel: visual sample of disturbed galaxies, 2nd panel down: cleaned-auto sample, 3rd panel down: union of visually and cleaned-auto-selected galaxies, bottom panel: intersection of visually and cleaned-auto-selected galaxies. \label{fig-density_radius_hist}}
\end{center}
\end{figure}

We quantify the local density using the standard measure of the projected
distance to the 10th nearest galaxy with $M_{r} < -18$ that is also within the
cluster's red sequence in an attempt to mitigate contamination. We find no
evidence that the rate of tidal features is dependent on local density as both
the visual and the cleaned-auto samples differ from the full cluster sample at
less than $2\sigma$ significance using a K-S test.  We next attempt to remove
the correlation between local density and clustercentric radius by  dividing the
galaxies into clustercentric annuli and within each annulus ranking galaxies
uniformly by local density to range from 1 to 10 (10 being most dense).  We then
sum the density ranks of disturbed galaxies for all distance bins and compare
that distribution to a random distribution (see Fig.
\ref{fig-density_radius_hist}).  By doing this, we are asking whether at each
radius the galaxies with tidal features are preferentially in environments of
either low or high local density. We find that disturbed galaxies have no
statistically significant deviations in ranks from a random distribution.

Although we find no evidence for local density having an effect on the incidence
of tidal features, we caution that this is simply an absence of evidence rather
than evidence of an absence, particularly because our measurement of local
density is so crude.

\section{Conclusions}

In a sample of 54 galaxy clusters $(0.04<z<0.15)$ containing 3551 early-type
galaxies suitable for study, we identify those with tidal features both
interactively and automatically. This constitutes the largest sample studied to
date for signs of environmental dependence in the incidence of tidal features.
We find tidal features in $\sim$ 3\% of galaxies in this sample, with data of
this particular depth.  Regardless of the method used to classify tidal
features, or the fidelity imposed on such classifications, we find a deficit of
tidally disturbed galaxies with decreasing clustercentric radius that is most
pronounced inside of $\sim 0.5R_{200}$. 

Although this trend could be attributed to a rise in galaxy-galaxy interactions
at large clustercentric distances, where the galaxy pair velocities might be
better tuned to mergers, an alternative interpretation is that tidal features
are preferentially erased at small clustercentric radii. We examine this
hypothesis with a simple toy model that links tidal feature survival time to the
local dynamical time and find that although qualitatively the effect is in the
correct sense to explain the data, the model falls somewhat short
quantitatively. Nevertheless, given the relative success and the limited
statistics, we cannot exclude this model as a possible explanation for the
radial trend in the incidence of tidal tails. 

We also search for a dependence of the incidence of tidal features on local
density and find no statistically significant evidence for such, even after
accounting for the correlation between clustercentric radius and local density.
Unfortunately, the large uncertainties in the measure of local environment limit
the interpretation of a null result.

We demonstrate that interesting behavior exists in the rate of tidal features
among cluster early-types as a function of clustercentric radius. This measure
should provide some guidance for models of merging among early-types, which is
conjectured to be significant in the build-up of the red sequence
\citep{bell,faber}.  Models of merging in and around clusters should aim to
provide merger estimates based on observable features rather than theoretical
constructs. We expect that such work will provide strong constraints on the
effect of the cluster environment on the structure of galaxy halos, the build-up
of the red sequence of galaxies, and the origin of the intracluster stellar
population.

\begin{acknowledgements} HH acknowledges support from Marie Curie IRG grant
230924 and the Netherlands Organisation for Scientific Research grant number
639.042.814.  This paper is based on observations obtained with
MegaPrime/MegaCam, a joint project of CFHT and CEA/DAPNIA, at the
Canada-France-Hawaii Telescope (CFHT) which is operated by the National Research
Council (NRC) of Canada, the Institute National des Sciences de l'Univers of the
Centre National de la Recherche Scientifique of France, and the University of
Hawaii. This work is based in part on data products produced at TERAPIX and the
Canadian Astronomy Data Centre as part of the Canada-France-Hawaii Telescope
Legacy Survey, a collaborative project of NRC and CNRS.  \end{acknowledgements}

\clearpage

\end{document}